\documentclass[onecolumn]{revtex4}
\usepackage{graphicx}
\usepackage{dcolumn}
\usepackage{color}
\usepackage{bm}
\usepackage{amsmath}
\usepackage{amsfonts}
\usepackage{amssymb}
\RequirePackage[colorlinks,citecolor=red,urlcolor=magenta,linkcolor=blue]{hyperref}
\def\eq#1{{Eq.~(\ref{#1})}}
\def\eqs#1{{Eqs.~(\ref{#1})}}
\def\EH{Einstein-Hilbert }
\def\LL{Lanczos-Lovelock }
\def\gr{general relativity}
\input epsf
\begin{document}

\title{Geometrical variables with direct thermodynamic significance in \LL gravity}

\author{Sumanta
Chakraborty \footnote{sumanta@iucaa.ernet.in;\\
sumantac.physics@gmail.com}}

\affiliation{IUCAA, Post Bag 4, Ganeshkhind,
Pune University Campus, Pune 411 007, India}

\author{T. Padmanabhan \footnote{paddy@iucaa.ernet.in}}

\affiliation{IUCAA, Post Bag 4, Ganeshkhind,
Pune University Campus, Pune 411 007, India}

\date{\today}
\begin{abstract}

It has been shown in an earlier work [arXiv:1303.1535] that there exists a pair of canonically conjugate variables $(f^{ab},N^a_{bc})$ in general relativity which also act as thermodynamically conjugate variables on any horizon. In particular their variations $(f^{bc}\delta N^a_{bc}, N^a_{bc}\delta f^{bc})$, which occur in the surface term of the \EH action, when integrated over a null surface, have direct correspondence with $(S\delta T,T\delta S)$ where $(T,S)$ are the temperature and entropy. We generalize these results to \LL\ models in this paper. We identify two such variables in \LL\ models such that (a)
our results reduce to that of 
\gr\ in the appropriate limit and (b) the
variation of the surface term in the action, when evaluated on a null surface, has 
direct thermodynamic interpretation as in the case of \gr. The variations again correspond to $S\delta T$ and $T\delta S$ where $S$ is now the appropriate Wald entropy for the \LL model. 
The implications are discussed. 

\end{abstract}
\maketitle

\section{Introduction}\label{Paper2:Sec:Intro}

Judicious use of the principle 
of equivalence and principle of general covariance leads to the conclusion that gravity can be described as arising from the curvature of spacetime. These principles also go a long way in determining the  kinematics of gravity viz. how spacetime curvature affects the dynamics of matter fields. 

The next step is to determine the dynamics of spacetime, viz. how the matter fields act as a source for curvature.  Unfortunately, there is no simple guiding principle to help us in this task. One possibility is to postulate that the field equations, derived from a suitable action principle, should not contain derivatives higher than second order. The most general action principle which will lead to this result is the \LL\ action in D dimensions \cite{Lanczos1932,Lovelock1971,Padmanabhan2010b,Padmanabhan2013b}. (In D=4, this reduces to standard \EH action.) This has been the conventional way of describing theories of gravity.

In recent years, investigations have revealed several intriguing features for this class of \LL models. They stem from the deep connection between gravitational dynamics and horizon thermodynamics, which we now know transcends general relativity.
These results  first emerged in the context of \gr\ from 
the seminal work by Bekenstein, Hawking, Davies, Unruh and others showing that horizons in general (and black holes in particular) possess thermodynamic attributes like entropy \cite{Bekenstein1973,Bekenstein1974} and temperature \cite{Hawking1975,Davies1976, Unruh1976,Hawking1977}. Since any suitable null surface can act as a local Rindler horizon for a class of observers, this allows one to introduce observer-dependent thermodynamic variables around any event in spacetime. Then it seems natural to think of spacetime as some kind of fluid with its thermodynamic properties arising from the dynamics of underlying ``atoms of spacetime." This emergent gravity paradigm (see, for a review \cite{Padmanabhan2005a,Padmanabhan2010a,Wald2001})
has received significant amount of support from later investigations, especially from the following results:
\begin{itemize}
 \item 
 The gravitational field equations reduce to simple thermodynamic identities 
on horizons for a wide class of gravity theories \textit{more general than} Einstein gravity 
\cite{Padmanabhan2002,Cai2005,Padmanabhan2006a,Akbar2006,Padmanabhan2006b,Padmanabhan2009}. The action functional for gravity can be expressed as a sum of a bulk term and a 
surface term with a ``holographic" relation between them. Again this result holds 
not only in Einstein gravity but also in a more general class of theories 
\cite{Padmanabhan2010b,Padmanabhan2006c,Kolekar2010,Kolekar2012b}. In fact, the action in all \LL models can be given a thermodynamic interpretation \cite{Kolekar2012b}.
\item
Gravitational field equations in all the \LL\ models can be derived from a thermodynamic 
extremum principle \cite{Padmanabhan2007,Padmanabhan2008} 
and the action functional itself can be given a thermodynamic interpretation \cite{Padmanabhan2005b,Padmanabhan2010e}.
\item
Gravitational field equation reduces to the Navier-Stokes equation of fluid 
dynamics in any spacetime, when  projected on an arbitrary null surface, thereby generalizing previous results 
for black hole spacetime \cite{Padmanabhan2011a,Kolekar2012a,Damour1982}.
\item
More recently \cite{Padmanabhan2013a} these ideas have been taken significantly further in the context of \gr\ and these results have also been generalized to \LL models \cite{Sumanta2014}. It has been demonstrated that the following results hold in all these theories:
(a) The total Noether charge in a 3-volume $\mathcal{R}$, related to the time evolution vector field, can be interpreted as the heat content of the boundary $\partial \mathcal{R}$ of the volume. This provides yet another holographic result connecting the bulk and boundary variables.
(b) The time evolution of the spacetime itself can be described in an elegant manner as being driven by  the departure from holographic equipartition, measured by  $(N_{\rm bulk} - N_{\rm sur})$. The metric will be time independent in the chosen foliation if  $N_{\rm sur} = N_{\rm bulk}$ which happens for all static geometries.
\end{itemize}
We thus have two possible routes towards gravitational dynamics. One is the conventional route, using the action functional and geometrical variables. The other is the thermodynamic route which uses suitably defined degrees of freedom, heat content etc. The link between these two routes, obviously, is provided by the action functional which -- as is known from previous investigations 
\cite{Padmanabhan2005b,Padmanabhan2010e,Padmanabhan2011a,Kolekar2012a,Kolekar2012b,
Padmanabhan2006c,Padmanabhan2010c} -- has both dynamical and thermodynamical interpretation. This is a peculiar aspect of gravitational action functionals, not shared by other theories which possess no thermodynamic or emergent interpretation and hence is worth probing deeply. 

A first step in this direction was taken in Ref. \cite{Krishna2013}. This work introduced two canonically conjugate variables
\begin{equation}
 f^{ab}=\sqrt{-g}g^{ab}; \qquad  N^{c}_{ab}=Q^{cp}_{aq}\Gamma ^{q}_{bp}+Q^{cp}_{bq}\Gamma ^{q}_{ap};
 \qquad Q^{ab}_{cd}=\frac{1}{2}\left(\delta ^{a}_{c}\delta ^{b}_{d}-\delta ^{a}_{d}\delta ^{b}_{c}\right)
\end{equation} 
in the context of general relativity and showed that (the variation of) these quantities also possess simple thermodynamical interpretation in terms of $S\delta T$ and $T\delta S$. But since the thermodynamic interpretation transcends \gr, we should be able to find similar geometrical variables in all \LL models. The main purpose of this paper is to identify one set of such variables which satisfy the conditions:
(a) These variables reduce to the ones used in \gr\ in $D=4$ when the \LL model reduces to \gr. (b) The variation of these quantities corresponds to $S\delta T$ and $T\delta S$ where $S$ is now the correct Wald entropy of the \LL model. 
We shall show that this can indeed be done and will discuss several properties of these variables.

The paper is organized as follows: In Sec. \ref{Paper2:Sec:LLIntro} 
we first review various properties of the \LL 
theories of gravity which we need for the rest of the paper. In the next section 
\ref{Paper2:Sec:ConjuLL} we summarize some of the results regarding 
canonically conjugate variables 
in \EH action and their possible generalization to \LL gravity. 
In Sec. \ref{Paper2:Sec:GNC} we present construction of 
the line element near an arbitrary null surface and 
the thermodynamic interpretation for conjugate variables 
related to \EH action. Section \ref{Paper2:Sec:ThermoLL} presents 
thermodynamic quantities for a general static spacetime and 
then it is generalized to arbitrary 
null surface constructed in the previous section for \LL gravity. The next 
section \ref{Paper2:Sec:GravLL} describes gravity in terms 
of the newly introduced conjugate variables 
in both \EH and \LL actions. 
Finally, we conclude with a discussion on our results. 
We have also presented some more details of the main calculations  in two Appendices, \ref{Paper2:AppA} and 
\ref{Paper2:AppB}.

The metric is assumed to have signature $(-,+,+,\ldots,+)$. 
The fundamental constants, $G$, $\hbar$ and $c$ are set 
to unity. The latin indices $a,b,\ldots$ run 
over all the spacetime coordinates, the greek indices $\mu,\nu, \ldots$ 
over the $(D-1)$ spatial coordinates and the capitalized 
latin indices $A,B,\ldots$ run over the $(D-2)$ transverse 
coordinates when relevant. 

\section{A Brief Introduction to \LL Gravity}\label{Paper2:Sec:LLIntro}

Consider a D-dimensional spacetime in which the gravity is 
described by an
action functional:
\begin{equation}\label{Paper2:Eq04}
A=\int _{\mathcal{V}}d^{D}x \sqrt{-g}L\left(g^{ab},R^{a}_{~bcd}\right)
\end{equation}
The Lagrangian depends both on the curvature and the metric but not on the derivatives of the curvature. 
The most important 
quantity for our later purpose, derived from the Lagrangian, is the tensor:
\begin{equation}\label{Paper2:Eq05}
P^{abcd}=\left(\frac{\partial L}{\partial R_{abcd}} \right)_{g_{ij}}
\end{equation}
having all the algebraic symmetry properties of the curvature 
tensor. An analogue of the Ricci tensor in \gr\ can also be constructed by the definition 
\begin{equation}\label{Paper2:Eq06}
\mathcal{R}^{ab}\equiv P^{aijk}R^{b}_{~ijk}.
\end{equation}
This tensor is indeed symmetric; but the result is nontrivial to prove 
(for this result and other properties of these tensors, see \cite{Padmanabhan2011a}).
The variation of the action presented in \eq{Paper2:Eq04} leads to the result:
\begin{eqnarray}\label{Paper2:Eq07}
\delta A&=&\delta \int _{\mathcal{V}} d^{D}x \sqrt{-g}L
\nonumber
\\
&=&\int _{\mathcal{V}}d^{D}x \sqrt{-g}E_{ab}\delta g^{ab}
+\int _{\mathcal{V}}d^{D}x \sqrt{-g}\nabla _{j}\delta v^{j}
\end{eqnarray}
where $E_{ab}$ is the field equation term and $\delta v^{a}$ is 
the boundary term. They are given by the following expressions:
\begin{eqnarray}\label{Paper2:Eq08}
E_{ab}&\equiv & \frac{1}{\sqrt{-g}}\left(\frac{\partial \sqrt{-g}L}{\partial g^{ab}} \right)_{R_{abcd}}
-2\nabla ^{m}\nabla ^{n}P_{amnb}
\nonumber
\\
&=&\mathcal{R}_{ab}-\frac{1}{2}g_{ab}L-2\nabla ^{m}\nabla ^{n}P_{amnb}
\\
\delta v^{j}&=&2P^{ibjd}\nabla _{b}\delta g_{di}-2\delta g_{di}\nabla _{c}P^{ijcd}
\end{eqnarray}
Since the quantity $P^{abcd}$ involves second order derivative of the metric, 
the term $\nabla ^{m}\nabla ^{n}P_{amnb}$ contains fourth order 
derivatives of the 
metric. Therefore, to get second order field equation, 
we must impose an extra condition on 
$P^{abcd}$, such that
\begin{equation}\label{Paper2:Eq09}
\nabla _{a}P^{abcd}=0.
\end{equation}
Thus the problem of finding an action functional leading to
a second order field equation  reduces to 
finding scalar functions of curvature and metric 
such that \eq{Paper2:Eq09} is satisfied. 
Such a scalar  indeed exists and is unique; it is  
given \cite{Lovelock1971,Padmanabhan2010b,Padmanabhan2013b,Padmanabhan2006a,Padmanabhan2006c} 
by the \LL Lagrangian in 
D dimensions, as
\begin{equation}\label{Paper2:Eq10}
L=\sum _{m}c_{m}L_{m}=\sum _{m}\frac{c_{m}}{m}
\frac{\partial L_{m}}{\partial R_{abcd}}R_{abcd}=\sum _{m}\frac{c_m}{m}P^{abcd}_{(m)}R_{abcd}
\end{equation}
This Lagrangian $L_{m}$ being a homogeneous function of 
$R_{abcd}$ of order $m$ 
can also be written as $L_{m}=Q^{abcd}_{(m)}R_{abcd}$, 
which can be used to identify 
$P^{abcd}_{(m)}=mQ^{abcd}_{(m)}$. 
From now on we shall work with this $m$th order Lagrangian 
and henceforth shall drop the $m$ index. For this 
Lagrangian we have the following explicit expression for $P_{cd}^{ab}$ 
in terms of the curvature tensor:
\begin{equation}\label{Paper2:Eq11}
P^{ab}_{cd}=\frac{\partial L_{m}}{\partial R^{cd}_{ab}}
=m\delta ^{aba_{2}b_{2}\ldots a_{m}b_{m}}_{cdc_{2}d_{2}\ldots c_{m}d_{m}}
R^{c_{2}d_{2}}_{a_{2}b_{2}}\ldots R^{c_{m}d_{m}}_{a_{m}b_{m}}\equiv mQ^{ab}_{cd}
\end{equation}
This relation will be used extensively later. Also note that due to complete antisymmetry 
of the determinant tensor in a D dimensional space-time we have 
the following restriction $2m\leq D$, otherwise the determinant tensor 
would vanish identically.
\LL models at dimension $D=2m$ are known as critical dimensions for a given \LL
term. In these critical dimensions the variation of the action functional reduces to a pure 
surface term \cite{Padmanabhan2011c}. 

All 
generally covariant theories, including the \LL theories of gravity, 
possess diffeomorphism invariance. This implies 
that the invariance under an infinitesimal coordinate transformation, $x^{a}\rightarrow x^{a}+\xi ^{a}(x)$ 
of the theory should lead to conservation of a current, usually called the 
\emph{Noether current}. From variation of 
action functional we can get the 
Noether current having the following expression 
\cite{Padmanabhan2013b,Padmanabhan2010b,Padmanabhan2011b,Padmanabhan2010d}:
\begin{equation}\label{Paper2:Eq12}
J^{a}\equiv \left( 2E^{ab}\xi _{b}+L\xi ^{a}+\delta _{\xi}v^{a}\right)
\end{equation}
In the above expression the last term, $\delta _{\xi}v^{a}$ represents the 
boundary term when the metric variation has the following form:
$\delta g^{ab}=\nabla ^{a}\xi ^{b}+\nabla ^{b}\xi ^{a}$. 
From the property of the 
Noether current $\nabla _{a}J^{a}=0$, 
we can define an antisymmetric tensor referred to 
as \emph{Noether Potential} by the condition, 
$J^{a}=\nabla _{b}J^{ab}$. Using 
\eq{Paper2:Eq08} we can substitute for 
the  the boundary term leading to an explicit form 
for both the Noether current and Potential. These general expressions can be found 
in \cite{Padmanabhan2010b}. In the context of
 \LL theories, where $\nabla _{a}P^{abcd}=0$ , they are given by
\begin{eqnarray}
J^{ab}&=&2P^{abcd}\nabla _{c}\xi _{d}
\label{Paper2:Eq13a}
\\
J^{a}&=&2P^{abcd}\nabla _{b}\nabla _{c}\xi _{d}
=2\mathcal{R}^{ab}\xi _{b}+2P_{k}^{~ija}\mathcal{L}_{\xi}\Gamma ^{k}_{ij}
\label{Paper2:Eq13b}
\end{eqnarray}
with $\Gamma ^{a}_{bc}$ being the metric compatible connection. 

The Noether current has a direct thermodynamic interpretation. To begin with one can associate a Wald entropy with horizons in all \LL models. The corresponding entropy density (which, integrated over the horizon gives the entropy) is given by \cite{Padmanabhan2010a,Padmanabhan2012b,Wald1993,
Strominger1996,Ashtekar1998,Bombelli1986,Padmanabhan2013c}
\begin{eqnarray}\label{Paper1:Eq55}
s=-\frac{1}{8}\sqrt{\sigma} P^{abcd}\mu _{ab}\mu _{cd}=\frac{1}{2}\sqrt{\sigma}
P^{\alpha b d\beta}r_{\alpha}u_{b}u_{d}r_{\beta}
\end{eqnarray}
It can be shown \cite{Padmanabhan2013a,Sumanta2014}
that  the Noether charge inside a bulk region is $\mathcal{R}$ equal to the heat content of the boundary surface $\partial \mathcal{R}$. That is,
\begin{eqnarray}\label{Paper1:Eq57}
\int _{\mathcal{V}}d^{D-1}x \sqrt{h} u^{a}J_{a}\left(\xi \right)
&=&\epsilon \int _{\partial \mathcal{V}}d^{D-2}x~T_{loc}s
\end{eqnarray}
which provides a direct thermodynamic as well as holographic interpretation of Noether charge.
(Here $\epsilon=\pm 1$ is appropriately chosen such the the outward normal is in the direction of the acceleration of the observers; for details of this result see \cite{Padmanabhan2013a,Sumanta2014}).

\section{Conjugate Variables in \EH action and Possible Generalization to \LL Gravity}
\label{Paper2:Sec:ConjuLL}

In this section we start by reviewing the conjugate variable structure in 
\EH action, which we will subsequently generalize to \LL gravity. We will 
first summarize the results for \gr\ and then attempt a  relatively simple generalization to \LL models, which, however, does not work. Taking a cue from this, we will define a  modified set of variables, which --- as we shall see --- work correctly. 

We start by reviewing the structural aspects of \EH action. 
In four dimensions, the standard \LL criteria 
uniquely identify the action functional to be \EH action:
\begin{equation}\label{Paper2:Eq01}
16\pi A_{EH}=\int _{\mathcal{V}}d^{4}x\sqrt{-g}L_{EH}=\int _{\mathcal{V}}d^{4}x\sqrt{-g}R
\end{equation}
Motivated by our desire to generalize the results to \LL models, we will rewrite the above action  in terms of the relevant $Q_{a}^{~bcd}$ for this $m=1$ \LL model so that the 
Lagrangian density $L_{EH}$ becomes:
\begin{equation}\label{Paper2:Eq02}
L_{EH}\equiv Q_{a}^{~bcd}R^{a}_{~bcd};~~~~
Q_{a}^{~bcd}=\frac{1}{2}\left(\delta ^{c}_{a}g^{bd}-\delta ^{d}_{a}g^{bc} \right)
\end{equation}
It is well known that the \EH  action is the sum of a
bulk term, which is quadratic in 
the derivatives of the metric and a surface term 
which contains all the second derivatives 
\cite{Eddington1924}. 
 This 
decomposition can be written as in terms of the 
variable $Q^{~bcd}_{a}$ as
\begin{eqnarray}\label{Paper2:Eq03}
L_{EH}&=&L_{quad}+L_{sur}
\nonumber
\\
L_{quad}&\equiv & 2Q_{a}^{~bcd}\Gamma ^{a}_{dk}\Gamma ^{k}_{bc};~~~
L_{sur}\equiv \frac{2}{\sqrt{-g}}\partial _{c}\left(\sqrt{-g}Q_{a}^{~bcd}\Gamma ^{a}_{bd}\right).
\end{eqnarray}
The bulk part of the action, when varied , leads 
to Einstein equation. 
We stress \cite{Padmanabhan2010b} the following point: \emph{The equations of motion can 
be determined from the the variation of the bulk term alone}. 
The solutions to the field equation 
(including, say, the black hole solutions) are not dependent on  the surface term in any way since 
they are derived from 
the bulk term alone. Nevertheless, when the surface term 
is evaluated on the horizon it produces 
the entropy associated with the horizon.
This is one of the hints for the thermodynamic interpretation arising from the action functional.

\subsection{Conjugate variables in Einstein gravity}\label{Paper2:Sec:ConjuLL:EH}

We now review the use of a different 
set of canonically conjugate variables in \EH action.
This  is done in detail in 
Ref. \cite{Padmanabhan2013a} and 
\cite{Krishna2013}; however for completeness, we review some of 
these results. 
The variational principle in \gr\
becomes simpler if we introduce a new dynamical 
variable $f^{ab}\equiv \sqrt{-g}g^{ab}$ (which is a tensor density) instead of the 
metric $g_{ab}$. Then 
the \EH action in terms of the new variable $f^{ab}$ 
reads \cite{Krishna2013,Padmanabhan2013a}
\begin{equation}\label{Paper2:Eq20}
\sqrt{-g}R=\sqrt{-g}L_{quad}-\partial _{c}\left[f^{ab}\frac{\partial 
\left(\sqrt{-g}L_{quad}\right)}{\partial\left(\partial _{c}f^{ab} \right)} \right]
=\sqrt{-g}L_{quad}-\partial _{c}\left(f^{ab}N^{c}_{ab}\right)
\end{equation}
where we have defined another quantity, representing the conjugate momenta to $f^{ab}$ as
\begin{equation}\label{Paper2:Eq21}
N^{c}_{ab}=\frac{\partial \left(\sqrt{-g}L_{quad}\right)}{\partial\left(\partial _{c}f^{ab} \right)} 
=Q_{ae}^{cd}\Gamma ^{e}_{bd}+Q_{be}^{cd}\Gamma ^{e}_{ad}.
\end{equation}
The usefulness of these variables is apparent from the structure of the variation of the \EH action. These 
variables also lead to two terms, the equation of motion term and the surface term having 
the following decomposition:
\begin{equation}\label{Paper2:Eq22}
\delta \left(\sqrt{-g}R\right)=R_{ab}\delta f^{ab}+f^{ab}\delta R_{ab}
=R_{ab}\delta f^{ab}-\partial _{c}\left(f^{ab}\delta N^{c}_{ab}\right).
\end{equation}
Making the surface term vanishing is equivalent to demanding  $\delta N^{a}_{bc}=0$, which is equivalent 
to setting variation of momentum to zero at the end points. 
(That is, we interpret \EH action as a momentum space action; 
see p.291 of ref. \cite{Padmanabhan2010b})
Also many other expressions 
simplifies considerably in terms of these variables \cite{Krishna2013,Padmanabhan2013a}. 
We summarize some of these important relations in terms of these variables below:
\begin{eqnarray}
R_{ab}&=&\left(-\partial _{c}N^{c}_{ab}-N^{c}_{ad}N^{d}_{bc}+\frac{1}{3}N^{c}_{ac}N^{d}_{bd}\right)
\label{Paper2:Eq23a}
\\
\sqrt{-g}R&=&-f^{ab}\partial _{c}N^{c}_{ab}-\frac{1}{2}N^{c}_{ab}\partial _{c}f^{ab}
\\
&=&\frac{1}{2}N^{c}_{ab}\partial _{c}f^{ab}-\partial _{c}\left(f^{ab} N^{c}_{ab}\right)
\label{Paper2:Eq23b}
\\
&=&-\frac{1}{2}\left[f^{ab}\partial _{c}N^{c}_{ab}+\partial _{c}\left(f^{ab}N^{c}_{ab}\right)\right]
\\
\sqrt{-g}L_{quad}&=&\frac{1}{2}N^{c}_{ab}\partial _{c}f^{ab};~~~
\sqrt{-g}L_{sur}=-\partial _{c}\left(f^{ab}N^{c}_{ab}\right)
\label{Paper2:Eq23c}
\end{eqnarray}
The Noether current also has a simple expression 
\cite{Krishna2013,Padmanabhan2013a} in terms of these variables:
\begin{equation}\label{Paper2:Eq24}
\sqrt{-g}J^{a}=2\sqrt{-g}R^{ab}\xi _{b}+f^{pq}\mathcal{L}_{\xi}N^{a}_{pq}
\end{equation}
Next we will consider the question of finding similar variables for \LL models.

\subsection{Conjugate variables in Lanczos-Lovelock theories}
\label{Paper2:Sec:ConjuLL:LL}

In Einstein gravity, we have obtained a simpler description 
using the two conjugate variables, $f^{ab}=\sqrt{-g}g^{ab}$ 
and $N^{c}_{ab}=Q_{ae}^{cd}\Gamma ^{e}_{bd}+Q_{be}^{cd}\Gamma ^{e}_{ad}$. 
The most natural choice for the corresponding  variables 
in \LL gravity will be the ones obtained by defining $N^{a}_{bc}$ by the same relation, viz. by \eq{Paper2:Eq21}, as in \gr\ but with $Q^{ab}_{cd}$ given by that for the appropriate \LL model, viz. by
\eq{Paper2:Eq11}. 
We will first explore this choice of  variable and  see 
whether they satisfy all the requirements.

Even in \LL theories, 
the decomposition of the Lagrangian into a bulk term and a 
surface term exists and has the identical expression as in 
\eq{Paper2:Eq03} with the $Q^{ab}_{cd}$ defined by \eq{Paper2:Eq11}.
There are two desirable features we expect the  variables to 
satisfy: First, the surface term should be expressible as, 
$-\partial (qp)$; second the quadratic part of the Lagrangian 
must be expressible as, $p\partial q$. Note that $q$ and $p$ 
are not absolute, since in the Hamiltonian formulation both 
are given equal weightage,and  we can even interchange $q$ and $p$. In 
\EH action these relations are given by \eq{Paper2:Eq23c}.

With this motivation, let us examine whether 
these two variables, $\tilde{f}^{ab}$ which is same as the $f^{ab}$ 
in \EH action and $\tilde{N}^{a}_{bc}$ having
the same expression as that of $N^{a}_{bc}$ with $Q^{ab}_{cd}$ 
corresponding to the one for the \LL model, satisfy our criteria. 
For that purpose we need to evaluate the following 
combinations, $\tilde{f}^{ab}\partial _{c}\tilde{N}^{c}_{ab}$ and 
$\tilde{N}^{a}_{bc}\partial _{a}\tilde{f}^{bc}$. These are evaluated explicitly in 
Appendix \ref{Paper2:AppA} [see \eq{Paper2:Eqa04} and \eq{Paper2:Eqa05}]. We  state here
the final results:
\begin{eqnarray}\label{Paper2:Eq25}
\sqrt{-g}L_{sur}&=&-\partial _{c}\left(\tilde{f}^{pq}\tilde{N}^{c}_{pq}\right)
\\
\sqrt{-g}L_{quad}&=&\tilde{N}^{c}_{ab}\partial _{c}\tilde{f}^{ab}
-\left(2\sqrt{-g}Q_{p}^{~bqc}\Gamma ^{p}_{qb}\Gamma ^{m}_{cm}
-2\sqrt{-g}g^{bm}Q_{ap}^{cq}\Gamma ^{p}_{qb}\Gamma ^{a}_{cm}\right)
\\
\sqrt{-g}Q_{p}^{~qrs}R^{p}_{~qrs}&=&-\tilde{f}^{ab}\partial _{c}\tilde{N}^{c}_{ab}
-\left(2\sqrt{-g}Q_{p}^{~bqc}\Gamma ^{p}_{qb}\Gamma ^{m}_{cm}
-2\sqrt{-g}g^{bm}Q_{ap}^{cq}\Gamma ^{p}_{qb}\Gamma ^{a}_{cm}\right)
\end{eqnarray}
Thus even though it leads to the proper surface term it does not simplify the other terms 
unlike in the \EH case presented in \eq{Paper2:Eq23c}. Hence we conclude that these 
variables are not suitable. 

The same conclusion can also be reached
from the  expression for the  Noether current as well. 
In \LL theories the Noether current has the expression given 
by \eq{Paper2:Eq13b}, which can actually be derived without 
any notion of diffeomorphism invariance, 
using only differential geometry \cite{Padmanabhan2013a,Sumanta2014}. 
As in the \EH scenario where the last term becomes, 
$f^{ab}\mathcal{L}_{\xi}N^{c}_{ab}$, here also we would like the last term to be of the 
form $\tilde{f}^{pq}\mathcal{L}_{\xi}\tilde{N}^{a}_{pq}$. 
However this does not yield the correct Noether current. 
An extra term, depending on Lie variation of the entropy 
tensor $P^{abcd}$, comes into the picture. We have used 
several identities regarding Lie variation of $P^{abcd}$ but 
none of these  help to simplify the extra term. [Though 
these identities do not help, they are quiet interesting and 
have not been derived earlier; hence  we present these 
relations in Appendix. \ref{Paper2:AppB}]. In summary, these variables, 
though they are the simplest  choice, do not fulfill 
the  criteria we would like them to satisfy. 

It is, however, possible to attack the problem from a 
different angle and obtain  another set of variables that satisfies 
all our criteria. The clue comes from the Noether current itself. We know from
\eq{Paper2:Eq24} that in \gr , the second term of the Noether current, given by 
$f^{ab}\mathcal{L}_{\xi}N^{c}_{ab}$,has the two variables $[f^{ab},N^{c}_{ij}]$
on the two sides of $\mathcal{L}_{\xi}$ while in the $m$th order \LL\ model the corresponding term  
$P_{k}^{~aji}\mathcal{L}_{\xi}\Gamma ^{k}_{aj}=
mQ_{k}^{~aji}\mathcal{L}_{\xi}\Gamma ^{k}_{aj}$ (see \eq{Paper2:Eq13b}) has the variables
$[\Gamma ^{k}_{aj},Q_{k}^{~aji}]$ on the two sides of $\mathcal{L}_{\xi}$. Taking a cue from this, let us define two variables: 
\begin{eqnarray}
\Gamma ^{a}_{bc}&=&\frac{1}{2}g^{ad}\left(-\partial _{d}g_{bc}
+\partial _{b}g_{cd}+\partial _{c}g_{bd}\right)
\label{Paper2:Eq26a}
\\
U_{a}^{~bcd}&=&2\sqrt{-g}Q_{a}^{~bcd}
\label{Paper2:Eq26b}
\end{eqnarray}
where  $\Gamma ^{a}_{bc}$, of course,  is the standard connection 
and $Q_{a}^{~bcd}$ is given by \eq{Paper2:Eq11}. (The factors are chosen to give the correct limit for \gr\ when $m=1$; we will see that these definitions work.)

Interestingly, these variables can be introduced in a somewhat different manner as well. Suppose we consider the \LL model Lagrangian which can be expressed entirely in terms of $R^{im}_{kl}$ [with index placements (2,2)] plus Kronecker deltas. Since $R^{im}_{kl}=g^{mj}R^i_{\phantom{i}jkl}$ it can also be expressed in terms of the variables $R^i_{\phantom{i}jkl}$ [with index placements (1,3)] and the metric $g^{ab}$. On the other hand,
$R^i_{\phantom{i}jkl}$ [with index placements (1,3)] can be written entirely in terms of $\partial_l\Gamma^i_{jk}$ and $\Gamma^i_{jk}$ without the metric appearing anywhere. Therefore, one can think of the \LL Lagrangian, when expressed in terms of $R^i_{\phantom{i}jkl}$ [with index placements (1,3)] and the metric $g^{ab}$ as a functional of $[g^{ab},\partial_l\Gamma^i_{jk},\Gamma^i_{jk}]$. This suggests defining the ``conjugate variable" to the connection: 
\begin{equation}
 mU_{u}^{~vlw}\equiv \frac{\partial \left(\sqrt{-g}L\right)}
{\partial \left(\partial _{l}\Gamma ^{u}_{vw}\right)}
\end{equation} 
The equality of the two sides, for $m-$th order \LL\ Lagrangian is easy to verify using \eq{Paper2:Eqanew}. It can be proved that this quantity $U_{abcd}$ has all the symmetries of the curvature tensor (essentially because the $\partial _{l}\Gamma ^{u}_{vw}$ occurs in $L$ only through the curvature tensor.)
This suggests yet another meaning to the  variables introduced  in \eqs{Paper2:Eq26a} 
and (\ref{Paper2:Eq26b}). 

Of course, this set of variables in  \eqs{Paper2:Eq26a} and (\ref{Paper2:Eq26b}) works for the \EH case well. Now  $U_{u}^{~vlw}$ can be expressed entirely in terms of the metric and -- in this sense -- one can think of the metric (or rather the particular combination of metric components) as conjugate to connection (rather than the other way around!) even in the \EH case.
All the original relations for the \EH action  can be written in 
terms of these two variables  instead of $f^{ab}$ and $N^{a}_{bc}$ 
as  in \eqs{Paper2:Eqa06} and (\ref{Paper2:Eqa07}) matching 
exactly \eq{Paper2:Eq23c}.
The reason has to do with the fact that for any covariant derivative operator $\hat{D}$ we have
\begin{eqnarray}\label{Paper2:Eq27}
f^{pq}\hat{D}N^{a}_{pq}&=&2\sqrt{-g}g^{pq} \hat{D} \left(Q^{ae}_{pd}\Gamma ^{d}_{eq}\right)
\nonumber
\\
&=&2\sqrt{-g}g^{pq}Q^{ae}_{pd}\hat{D}\Gamma ^{d}_{eq}
\nonumber
\\
&=&2\sqrt{-g}Q_{d}^{~qea}\hat{D}\Gamma ^{d}_{qe}=U_{p}^{~qra}\hat{D}\Gamma ^{p}_{qr}.
\end{eqnarray}
This is possible only in the \EH action where $Q_{ab}^{cd}$ involves only 
the Kronecker delta functions and hence can be pulled through any 
covariant derivative operator. Thus, 
in general, the pair  [$\Gamma ^{a}_{bc}, U_{a}^{~pqr}$] can also act as 
the conjugate variables, in Einstein gravity. Since $U_{a}^{~pqr}$ can be entirely expressed in terms of the metric in \gr\ , they
are equivalent to $f^{ab}$ and $N^{c}_{ab}$ structurally. 
However in higher order \LL theories 
$Q^{ab}_{cd}$  depends on curvature tensor and hence the 
above equivalence is broken. As we shall see it is better to work with the two variables as  in 
\eqs{Paper2:Eq26a} and (\ref{Paper2:Eq26b}), as shown in App. \ref{Paper2:AppA} 
\eqs{Paper2:Eqa06} and (\ref{Paper2:Eqa07}).

Finally, note that, by construction, the Noether current  becomes in terms of these variables
\begin{eqnarray}\label{Paper2:Eq28}
\sqrt{-g}J^{a}&=&m\left(2U_{b}^{~pqr}R^{a}_{~pqr}v^{b}+
U_{b}^{~cda}\mathcal{L}_{\xi}\Gamma ^{b}_{cd}\right)
\nonumber
\\
&=&2\sqrt{-g}\mathcal{R}^{a}_{b}v^{b}+mU_{b}^{~cda}\mathcal{L}_{\xi}\Gamma ^{b}_{cd}
\end{eqnarray} 
This is exactly of the same form as \eq{Paper2:Eq24}, the Noether current 
for \EH action.

\section{Thermodynamic structure of 
Einstein-Hilbert action}\label{Paper2:Sec:GNC}

In this section we will rapidly review 
the thermodynamic structure of conjugate variables 
introduced in an earlier work \cite{Krishna2013} and then 
shall extend these thermodynamic variables to \LL gravity.  
For the geometrical construction, we shall confine ourselves 
in a four-dimensional spacetime, 
with the possibility of generalizing to higher 
dimensions in a straightforward manner. 
We start by reviewing the construction 
of metric near an arbitrary null surface.

\subsection{Construction of Gaussian null coordinates}
\label{Paper2:Sec:GNC:Const}

Let us consider the four-dimensional spacetime $V^{4}=M^{3}\times R$, where $M^{3}$ is a compact 
three-dimensional manifold. We will consider spacetimes to be time orientable with null embedded 
hypersurfaces, which are diffeomorphic to $M^{3}$ with closed null generators. 
We take $\mathcal{N}$ to be such a null hypersurface with 
 null generator \cite{Moncrief1983,Morales2008}. 
On this null surface $\mathcal{N}$ we can introduce spacelike two surface with coordinates 
$(x_{1},x_{2})$ defined on them. The null geodesics generating the null hypersurface 
$\mathcal{N}$ goes out of this  spacelike two surface. Thus we can use these null generators 
to define coordinates on the null hypersurface. The intersection point of these null geodesics 
with the spacelike two surface can be determined by the coordinates $(x_{1},x_{2})$, which 
then evolves along the null geodesic, which is affinely parametrized by $u$, and label each point 
on the null hypersurface as $(u,x_{1},x_{2})$. The above system of coordinates readily 
identify three basis vectors: (a) the tangent to the null geodesics, 
${\bm \ell}=\partial /\partial u$, and (b) basis vectors on the two surface 
$\textbf{e}_{A}=\partial /\partial _{A}$.

Having fixed the coordinates on the null surface $\mathcal{N}$ we now move out of the 
surface using another set of null generators with tangent $k^{a}$ satisfying the
following constraints: (a) $k_{a}k^{a}=0$, (b) $e^{a}_{A}k_{a}=0$ and finally $\ell _{a}k^{a}=-1$. 
These null geodesics are taken to intersect the null surface at coordinates 
$(u,x_{1},x_{2})$ and then move out with affine parameter $r$, such that any point in the 
neighborhood of the null surface can be characterized by four coordinates $(u,r,x_{1},x_{2})$. 
In this coordinate system, the null surface is given by the condition $r=0$. 
This defines a coordinate system 
$\lbrace u,r,x^{A}\rbrace$ over the global manifold $V^{4}$. This system of coordinates are
formed in a manner analogous with Gaussian Normal Coordinate and hence is 
referred to as \emph{Gaussian null coordinates} (\emph{GNC}).

Having set the full coordinate map near the null surface we now proceed to determine the 
metric elements in that region. 
Note that $\ell _{a}\ell ^{a}=0$ leads to $g_{uu}=0$ on the null surface since 
${\bm \ell}=\partial /\partial u$. We also note that the basis vectors $e^{a}_{A}$ 
have to lie on the null surface implying $\ell _{a}e^{a}_{A}=0$ on $\mathcal{N}$, which leads 
to $g_{uA}=0$. Also the metric on the two-surface is given by $g_{AB}=e^{a}_{A}e^{b}_{B}g_{ab}$, 
which we denote by $\mu _{AB}$. We also need the criteria that $\mu _{AB}$ is positive 
definite with finite determinant ensuring invertibility and non-degeneracy of the two metric. 
Thus the following metric components gets fixed to be:
\begin{eqnarray}\label{Paper2:Eq29}
g_{uu}|_{r=0}=g_{uA}|_{r=0}=0;
\nonumber
\\
g_{AB}=\mu _{AB}
\end{eqnarray}
Let us now proceed to determine the other components of the metric. For this, we will 
use the vector $\mathbf{k}=-\partial /\partial r$ such that from $k^{a}k_{a}=0$ we get 
$g_{rr}=0$ throughout the spacetime manifold. Also from the criteria that the null geodesics 
are affinely parametrized by $r$ we readily obtain $\partial _{r}g_{r\alpha}=0$, where 
$\alpha =(u,x_{1},x_{2})$. Again, from the conditions $\ell ^{a}k_{a}=-1$, we readily get 
$g_{ru}=1$ and from $k_{a}e^{a}_{A}=0$ we get $g_{rA}=0$. From the criteria derived earlier 
showing $\partial _{r}g_{r\alpha}=0$ we can conclude that the above two-metric coefficients are 
valid everywhere. Thus within the global region $V^{4}$ we can have smooth 
functions $\alpha$ and $\beta _{A}$ such that, 
$\alpha \mid_{r=0}=\left(\partial g_{uu}/\partial r \right)\mid_{r=0}$ and 
$\beta _{A}\mid _{r=0}=\left(\partial g_{uA}/\partial r \right)\mid _{r=0}$. 
With these two 
identifications we have the following expression for the line element as
\begin{equation}\label{Paper2:Eq30}
ds^{2}=g_{ab}dx^{a}dx^{b}=-2r\alpha du^{2}+2dudr-2r\beta _{A}dudx^{A}+\mu _{AB}dx^{A}dx^{B}
\end{equation}
where $\mu _{AB}$ is the two-dimensional metric representing the metric on the 
null surface. Note that the construction presented above is completely general; it can 
be applied in the neighborhood of any null hypersurface and, in particular, to the event horizon of 
a black hole. 

\subsection{Thermodynamics related to Einstein-Hilbert action}
\label{Paper2:Sec:GNC:EHThermo}

In this section we will show that the variables $f^{ab}=\sqrt{-g}g^{ab}$ 
and $N^{a}_{bc}=Q^{ad}_{be}\Gamma ^{e}_{cd}+Q^{ad}_{ce}\Gamma ^{e}_{bd}$ are closely related to the thermodynamic properties of null surfaces. The variations $p\delta q$ and $q \delta p$ obtained from 
conjugate variables have direct thermodynamic interpretation associated with them. This result was obtained earlier in Ref. \cite{Krishna2013} but we go through the key steps for completeness as well as for a minor relaxation of the conditions originally used in Ref. \cite{Krishna2013}.
 
\subsubsection{Static Spacetime}\label{Paper2:Sec:GNC:EHThermo:Stat}

We will first prove these relations for a general static spacetime before discussing arbitrary 
null surface constructed above. 
Let us consider an arbitrary static spacetime with horizon. For this 
spacetime we have an arbitrary two surface 
with metric $\sigma _{AB}$ and the line 
element can be written in the form \cite{Medved2004}:
\begin{equation}\label{Paper2:Eq31}
ds^{2}=-N^{2}dt^{2}+dn^{2}+\sigma _{AB}dy^{A}dy^{B}
\end{equation}
In the above line element, $n$ represents spatial direction normal to the $(D-2)$ dimensional 
hypersurface with $\sigma _{AB}$ being transverse metric on the two surface. 
Let ${\bm \xi} =\partial /\partial t$ be a timelike Killing vector, with 
Killing horizon located at, $N^{2}\rightarrow 0$. The coordinate system is chosen in such a way that 
$n=0$ on the Killing horizon. Then, in the near horizon regime, the following expansions of 
$N$ and $\sigma _{AB}$ are valid \cite{Medved2004}:
\begin{eqnarray}\label{Paper2:Eq32}
N&=&\kappa (x^{A}) n+\mathcal{O}(n^{3})
\nonumber
\\
\sigma _{AB}&=&\left[\sigma _{H}(y)\right]_{AB}+\frac{1}{2}\left[\sigma _{2}(y)\right]_{AB}n^{2}+\mathcal{O}(n^{3})
\end{eqnarray}
In the above expression $\kappa$ is the local gravitational acceleration defined as: 
$\kappa =\partial _{n}N$ and in the $r\rightarrow 0$ limit, 
$\kappa \rightarrow \kappa _{H}$, satisfying all the standard properties 
of surface gravity. Also $\kappa /N$ represents the normal component 
of the four acceleration of an observer at fixed $(n,x^{A})$. 
Throughout the calculation we shall evaluate quantities on a 
$n=\textrm{constant}$ surface and then take the 
$n\rightarrow 0$ limit. The nonzero components 
of the metric variations are:
\begin{equation}\label{Paper2:Eq33}
\delta g_{tt}=2N\delta N; ~~~\delta g_{AB}=\delta \sigma _{AB}
\end{equation}
We will also require the nonzero components of $N^{a}_{bc}$ which are \cite{Krishna2013}:
\begin{equation}\label{Paper2:Eq34}
N^{n}_{tt}=-N\partial _{n}N;~~~N^{n}_{nn}=\frac{\partial _{n}N}{N}
+\frac{1}{2}\sigma ^{AB}\partial _{n}\sigma _{AB};~~~N^{n}_{AB}=\frac{1}{2}\partial _{n}\sigma _{AB}
\end{equation}
Then in the near horizon limit the respective $p\delta q$ and $q\delta p$ terms turn out to be:
\begin{equation}\label{Paper2:Eq35}
N^{n}_{ab}\delta f^{ab}|_{H}=2\kappa \delta (\sqrt{\sigma});
~~~f^{ab}\delta N^{n}_{ab}=2\sqrt{\sigma}\delta \kappa
\end{equation}
Then integration over the transverse variables and introduction of proper numerical 
factor yields the variation of the surface term in \EH action to be :
\begin{eqnarray}
\frac{1}{16\pi}\int d^{2}x_{\perp}N^{n}_{ab}\delta f^{ab}
&=&\int d^{2}x\frac{\kappa}{2\pi}\delta \left(\frac{\sqrt{\sigma}}{4}\right)=\int d^{2}xT\delta s
\label{Paper2:Eq36a}
\\
\frac{1}{16\pi}\int d^{2}x_{\perp} f^{ab}\delta N^{n}_{ab}
&=&\int d^{2}x\frac{\sqrt{\sigma}}{4}\delta \left(\frac{\kappa}{2\pi} \right)=\int d^{2}xs\delta T
\label{Paper2:Eq36b}
\end{eqnarray}
where $s=\sqrt{\sigma}/4$ is the entropy density of the spacetime. 
(If  the surface gravity $\kappa$ 
is independent of the transverse coordinates then the above results become $T\delta S$ and $S\delta T$ 
respectively, but the result in the above form expressed through 
\eqs{Paper2:Eq36a} and (\ref{Paper2:Eq36b}) is more generally applicable.) The above equations can also be written in terms of $U_{a}^{~bcd}$ 
and $\Gamma ^{a}_{bc}$, since
$f^{ab}\delta N^{c}_{ab}=2\sqrt{-g}g^{ab}Q^{ce}_{ad}\delta \Gamma ^{d}_{be}
=U_{d}^{~bec}\delta \Gamma ^{d}_{be}$
Hence \eqs{Paper2:Eq36a} and (\ref{Paper2:Eq36b}) 
can also be presented in terms of the variables $U_{a}^{~bcd}$ 
and $\Gamma ^{a}_{bc}$ as:
\begin{eqnarray}
\delta H_{sur}^{(1)}&=&\frac{1}{16\pi}\int d^{2}x~U_{a}^{bcr}\delta \Gamma ^{a}_{bc}
=\int d^{2}x~s\delta T
\label{Paper2:Eq37a}
\\
\delta H_{sur}^{(2)}&=&\frac{1}{16\pi}\int d^{2}x~\Gamma ^{a}_{bc}\delta U_{a}^{bcr}
=\int d^{2}x~T\delta s
\label{Paper2:Eq37b}
\end{eqnarray}  
However, as we will see later, this not possible in \LL gravity, 
and we have to work with $U_{d}^{~bec}$ directly.

\subsubsection{Generalization to Arbitrary null surface}
\label{Paper2:Sec:GNC:EHThermo:Null}

In the context of horizon thermodynamics the concepts 
of local Rindler frame and Rindler horizons are extensively 
used \cite{Padmanabhan2010c}, where local Rindler horizon refers 
to a patch of null surface which is perceived as the horizon by a 
suitably defined accelerated observer. This analysis can be made more general 
by working in the near horizon regime of an arbitrary null surface. Its  construction was 
presented in Sec. \ref{Paper2:Sec:GNC:Const} leading to the line element in 
the near horizon limit being given by \eq{Paper2:Eq30}, with $r=0$ representing the null 
surface. 

It will, however, be unrealistic to expect \textit{arbitrary} variations of 
the metric in the  near horizon 
limit of an arbitrary null surface to have simple thermodynamic interpretation. In order to 
obtain suitable thermodynamic interpretation we need to impose some restrictions. These restriction 
can be of two types: (i) Those involving restriction on the background 
metric itself with variations being arbitrary. This is broadly what we did in the last section where we assumed the background to be static but kept the variations arbitrary. (ii) Those keeping background metric arbitrary 
with some restrictions being imposed on the  variations. Among these two, restriction on the variation is conceptually preferable 
since we want the background metric to represent any 
\textit{arbitrary} null surface and hence we shall keep it most general. As we shall see, the conditions on the metric 
variation we need to impose turn out to be $\partial _{u}\delta (\mu _{AB})=0$ 
and  $\partial _{u}\delta (\sqrt{\mu})=0$. These conditions imply that these specific variations 
should not depend on time coordinate $u$ and thus henceforth will be referred to as stationarity 
conditions in what follows. (These conditions also generalize slightly the results obtained earlier in \cite{Krishna2013}).

We now proceed to calculate the surface term which turns out to be:
\begin{equation}
A_{sur}=-\frac{1}{16\pi}\int d^{3}x ~n_{c}f^{ab}N^{c}_{ab}
\end{equation}
One of the integral in the above expression is over time which will just lead to a multiplicative factor in \textit{static} geometries. It is, therefore,  more convenient to work with the 
surface Hamiltonian (which is actually
the heat density of the surface; see \cite{Padmanabhan2013a}) defined as:
\begin{equation}
H_{sur}=-\frac{\partial A_{sur}}{\partial u}=\frac{1}{16\pi} \int d^{2}x~n_{c} f^{ab}N^{c}_{ab}
\end{equation}
The surface Hamiltonian in \EH action, evaluated for the 
above metric, turns out to be \cite{Krishna2013}:
\begin{eqnarray}\label{Paper2:Eq41}
H_{sur}&=&-\frac{1}{16\pi}\int d^{2}x \sqrt{h}n_{r}V^{r}
\nonumber
\\
&=&-\frac{1}{16\pi}\int d^{2}x \Big[ -\frac{1}{\sqrt{\mu}}\partial _{u}\mu -\sqrt{\mu}
\left(2\alpha +2r\partial _{r}\alpha +2r\beta ^{2}+2r^{2}\beta _{A}\partial _{r}\beta ^{A}\right)
\nonumber
\\
&-&\frac{2r\alpha +r^{2}\beta ^{2}}{\sqrt{\mu}}\partial _{r}\mu -\sqrt{\mu}r\partial _{A}\beta ^{A}
-\frac{r\beta ^{A}}{\sqrt{\mu}}\partial _{A}\mu \Big].
\end{eqnarray}
Since integration variables are $u$ and $x^{A}$, we can take the $r\rightarrow 0$ limit easily 
leading to, $\sqrt{h}n_{r}V^{r}=-\partial _{u}\mu /\sqrt{\mu}-2\sqrt{\mu}\alpha$.  
The first term vanishes when stationarity conditions are used. We then  arrive at:
\begin{eqnarray}
H_{sur}=\frac{1}{16\pi}\int d^{2}x~2\sqrt{\mu}\alpha =\int d^{2}x \left(\frac{\alpha}{2\pi}\right)
\left(\frac{\sqrt{\mu}}{4}\right)=\int d^{2}x Ts
\end{eqnarray}
which is indeed the heat density at the boundary.

The variation of this surface Hamiltonian splits into two parts such 
that:
\begin{equation}\label{Paper2:Eq38}
\delta H_{sur}=\frac{1}{16\pi}\int d^{2}x~n_{c}\left(f^{ab}\delta N^{c}_{ab}
+N^{c}_{ab}\delta f^{ab}\right)
\equiv \delta H_{sur}^{(1)}+\delta H_{sur}^{(2)}
\end{equation}
where $\delta H_{sur}^{(1)}$ contains $f^{ab}\delta N^{c}_{ab}$ and 
$\delta H_{sur}^{(2)}$ 
contains $N^{c}_{ab}\delta f^{ab}$. Among 
them $f^{ab}\delta N^{c}_{ab}$ has the following expression near the 
$r=0$ null surface:
\begin{equation}\label{Paper2:Eq39}
f^{ab}\delta N^{r}_{ab}=2\sqrt{\mu}\delta \left(\alpha +\frac{1}{4}\mu ^{AC}\partial _{u}\mu _{AC}\right)
+\frac{1}{2}\sqrt{\mu}\mu ^{AB}\partial _{u}\delta \mu _{AB}
\end{equation}
It is evident from the above expression that the stationarity conditions mentioned above leads to
the following result:
\begin{equation}\label{Paper2:Eq40}
\delta H_{sur}^{(1)}\equiv
\frac{1}{16\pi}\int d^{2}xf^{ab}\delta N^{r}_{ab}=\frac{1}{8\pi}\int d^{2}x\sqrt{\mu}\delta \alpha
\end{equation}
Rather than evaluating the second term $N^{c}_{ab}\delta f^{ab}$ individually 
it is a little simpler to obtain it by considering variation 
of the total surface Hamiltonian. 
The variation of the full surface term is:
\begin{equation}\label{Paper2:Eq42}
\delta H_{sur}=\frac{1}{16\pi}\int d^{2}x\delta 
\left(2\sqrt{\mu}\alpha\right).
\end{equation}
Using this result and \eq{Paper2:Eq40} the second variation 
in \eq{Paper2:Eq38} becomes,
\begin{equation}\label{Paper2:Eq43}
\delta H_{sur}^{(2)}=
\frac{1}{16\pi}\int d^{2}xN^{r}_{ab}\delta f^{ab}=\frac{1}{8\pi}\int d^{2}x \alpha \delta \sqrt{\mu}
\end{equation}
We can rewrite the integrands of \eqs{Paper2:Eq40} and (\ref{Paper2:Eq43}) 
as $s\delta T$ and $T\delta s$ respectively. 
More formally, we write:
\begin{eqnarray}
\delta H_{sur}^{(1)}&=&\frac{1}{16\pi}\int d^{2}x f^{ab}\delta N^{r}_{ab}
\nonumber
\\
&=&\int d^{2}x\frac{\sqrt{\mu}}{4}\delta \left(\frac{\alpha}{2\pi}\right)
=\int d^{2}x~s\delta T
\label{Paper2:Eq44a}
\\
\delta H_{sur}^{(2)}&=&\frac{1}{16\pi}\int d^{2}x N^{r}_{ab} \delta f^{ab} 
\nonumber
\\
&=&\int d^{2}x\frac{\alpha}{2\pi}\delta \left(\frac{\sqrt{\mu}}{4}\right)
=\int d^{2}x~T\delta s
\label{Paper2:Eq44b}
\end{eqnarray}
where again $s=\sqrt{\mu}/4$ is the entropy density of the spacetime. 

 This result brings out the connection between these conjugate variables and respective thermodynamic 
quantities pertaining to the null surface acting as a local Rindler horizon. The curious feature 
is that while the surface term integrated  over the horizon leads to the heat content, 
the variation $T\delta s$ itself comes from variation of generalized coordinates and the variation 
$s\delta T$ coming from variation of generalized momenta. 

Finally, as in the previous section these relations can also be expressed 
in terms of $U_{a}^{~bcd}\delta \Gamma ^{a}_{bc}$ 
and $\Gamma ^{a}_{bc}\delta U_{a}^{~bcd}$ respectively. That is
\begin{eqnarray}
\delta H_{sur}^{(1)}&=&\frac{1}{16\pi}\int d^{2}x~U_{a}^{bcr}\delta \Gamma ^{a}_{bc}
=\int d^{2}x~s\delta T
\\
\delta H_{sur}^{(2)}&=&\frac{1}{16\pi}\int d^{2}x~\Gamma ^{a}_{bc}\delta U_{a}^{bcr}
=\int d^{2}x~T\delta s
\end{eqnarray}  
Next we will take up the case of \LL models.

\section{Thermodynamics Related to \LL Action}\label{Paper2:Sec:ThermoLL}

In this section, we shall describe the corresponding results 
for the \LL models. As in the previous case,  we 
shall first illustrate the results for a general static spacetime 
and then  generalize these results to 
the arbitrary null surface constructed in Sec. \ref{Paper2:Sec:GNC:Const}. 

The surface term in \LL Lagrangian, 
discussed in Sec. \ref{Paper2:Sec:LLIntro},  can be written as:
\begin{equation}\label{Paper2:Eq46}
-\partial _{c}\left(2\sqrt{-g}Q_{a}^{~bdc}\Gamma ^{a}_{bd}\right)\equiv 
-\partial _{c}\left(\sqrt{-g}V^{c}\right)
\end{equation}
Then under infinitesimal variation the surface term variation can be subdivided into two parts
\begin{equation}\label{Paper2:Eq47}
\delta \left(\sqrt{-g}V^{c}\right)=U_{a}^{~bdc}\delta \Gamma ^{a}_{bd}+\Gamma ^{a}_{bd}\delta U_{a}^{~bdc}
\end{equation}
where one term involves variation of connections, 
while the other one is quiet complex and involves variation of both
the metric and the entropy tensor. Hence the variation of the surface term can 
be written by introducing the surface Hamiltonian in an identical 
manner as follows:
\begin{eqnarray}\label{Paper2:Eq48}
\delta H_{sur}&=&\frac{1}{16\pi}\int d^{D-2}x~n_{c}
\delta \left(\sqrt{-g}V^{c}\right)
\nonumber
\\
&=&\frac{1}{16\pi}\int d^{D-2}x~n_{c}U_{a}^{~bdc}\delta \Gamma ^{a}_{bd}
+\frac{1}{16\pi}\int d^{D-2}x~n_{c}\Gamma ^{a}_{bd}\delta U_{a}^{~bdc}
\nonumber
\\
&=&\delta H_{sur}^{(1)}+\delta H_{sur}^{(2)}
\end{eqnarray}
Also the variation of the connection due to infinitesimal change of metric 
$g_{ab}\rightarrow g_{ab}+h_{ab}$ is:
\begin{equation}\label{Paper2:Eq49}
\delta \Gamma ^{p}_{qr}=\frac{1}{2}h^{pa}\left(-\partial _{a}g_{qr}
+\partial _{q}g_{ar}+\partial _{r}g_{aq}\right)
+\frac{1}{2}g^{pa}\left(-\partial _{a}h_{qr}+\partial _{q}h_{ar}+\partial _{r}h_{aq}\right)
\end{equation}
Using these results we can calculate the variation explicitly 
for different metrics. We will show that, just as in the case of \EH action,  the first term 
in \eq{Paper2:Eq48} leads to $s\delta T$ while the second term leads to 
$T\delta s$. 

\subsection{A general static spacetime}

We will consider an arbitrary static spacetime with horizons as presented 
in Sec. \ref{Paper2:Sec:GNC:EHThermo:Stat}. For the line element presented 
in \eq{Paper2:Eq31} using \eq{Paper2:Eqa08} and 
\eq{Paper2:Eq32} we find that  (in the relevant $n\rightarrow 0$ 
limit)  only three components of connection are non zero. These are : 
$\Gamma ^{n}_{tt}=N\partial _{n}N$, 
$\Gamma ^{t}_{nt}=\partial _{n}N/N$ and $\Gamma ^{A}_{BC}$. 
Another two expressions  we need for the calculation are the $(D-2)$ dimensional surface element 
and Wald entropy for \LL theories. The null surface we are interested in can be defined 
by the condition, $l^{2}=0$, with $l_{a}$ being (local) Killing vector. Then if we 
can introduce another auxiliary null vector $k_{a}$ such that, $l_{a}k^{a}=-1$, then the $(D-2)$ 
dimensional surface element turns out to be:
\begin{equation}\label{Paper2:Eq50}
d\Sigma _{ab}=d^{D-2}x\mu _{ab}=-d^{D-2}x\left(l_{a}k_{b}-l_{b}k_{a}\right).
\end{equation}
The Wald entropy expressed in terms of the entropy tensor is 
\cite{Wald1993,Padmanabhan2012a}:
\begin{equation}\label{Paper2:Eq51}
S=-\frac{1}{8}\int \sqrt{\sigma} d^{D-2}x P^{abcd}\mu _{ab}\mu _{cd}
\end{equation}
where $\sigma$ is the determinant of the metric on the two surface.
With  the $l_{a}$ and $k_{a}$ defining the bi-normal
of the $(D-2)$ dimensional surface, we have the following expression for the entropy of this static horizon:
\begin{equation}\label{Paper2:Eq52}
S=\frac{1}{2}\int d^{D-2}x\sqrt{\sigma} P^{nt}_{nt}\equiv \int d^{D-2}x s
\end{equation}
where $s=(1/2)\sqrt{\sigma} P^{nt}_{nt}$. 
Next we evaluate the variation $U_{a}^{~bdn}\delta \Gamma ^{a}_{bd}$ which turns out to be:
\begin{eqnarray}\label{Paper2:Eq53}
U_{a}^{~bdn}\delta \Gamma ^{a}_{bd}\mid _{H}&=&2\sqrt{-g}Q_{a}^{~bdn}\delta \Gamma ^{a}_{bd} 
\nonumber
\\
&=&2N^{2}\sqrt{\sigma}Q_{n}^{~ttn}\delta \left(\partial _{n}N\right)+2\sqrt{\sigma}Q_{t}^{~ntn}\delta \left(\partial _{n}N\right)
\nonumber
\\
&=&4\sqrt{\sigma}Q_{nt}^{nt}\delta \kappa
\end{eqnarray}
Along the similar lines we can compute the other part of the variation, $\Gamma ^{a}_{bd}\delta U_{a}^{~bdn}$ which can be written 
as:
\begin{equation}\label{Paper2:Eq54}
\Gamma ^{a}_{bd}\delta U_{a}^{~bdn}\mid _{H}=4\kappa \delta \left(\sqrt{\sigma}Q_{nt}^{nt}\right)
\end{equation}
From \eq{Paper2:Eq52} we get the expression for entropy and the temperature is given by 
as $\kappa /2\pi$. With these two identifications we find:
\begin{eqnarray}
\frac{1}{16\pi}\int d^{D-2}x~\Gamma ^{a}_{bd}\delta (mU_{a}^{~bdn})&=& 
\frac{1}{2}\int d^{D-2}x~\frac{\kappa}{2\pi}~\delta\left(\sqrt{\sigma} P^{nt}_{nt} \right)=\int d^{D-2}x T\delta s
\label{Paper2:Eq55a}
\\
\frac{1}{16\pi}\int d^{D-2}x~(mU_{a}^{~bdn})\delta \Gamma ^{a}_{bd}&=&
\frac{1}{2}\int d^{D-2}x\sqrt{\sigma} P^{nt}_{nt}
\delta \left(\frac{\kappa}{2\pi}\right)=\int d^{D-2}x~s\delta T
\label{Paper2:Eq55b}
\end{eqnarray}
In the above expressions $s$ represents the entropy density of the 
horizon, which reduces to 
$\sqrt{\sigma}/4$ in the \EH limit. The above results show that the two terms 
in \eq{Paper2:Eq48} leads to $T\delta s$ and $s\delta T$ 
respectively, with $\kappa$ being $\kappa (x^{A})$. Also we observe that these 
are identical to those obtained in \EH scenario as presented in \eqs{Paper2:Eq37a} 
and (\ref{Paper2:Eq37b}) respectively.
(As in the \gr, if $\kappa$ is independent of transverse coordinates 
\cite{Racz1996,Carter2010} one can integrate 
to obtain $T\delta S$ and $S \delta T$ respectively; but these results are more general.) 

Thus we have natural generalization of the results in 
\gr\ to all \LL models if we use the variables 
$U_{a}^{bcd}$ and $\Gamma ^{a}_{bc}$ introduced in 
Sec. \ref{Paper2:Sec:ConjuLL}.

\subsection{Generalization to arbitrary null surface}

Having discussed the thermodynamic interpretation 
of the two variables for 
a general static spacetime, we will now 
extend the result  to an arbitrary null surface. 
The metric near the null surface has been constructed in 
Sec. \ref{Paper2:Sec:GNC:Const} and we shall 
 use that metric to evaluate various quantities of 
interest. As in the case of static situation, here also 
we shall calculate all the quantities on a 
$r=\textrm{constant}$ surface and then shall take the limit 
$r\rightarrow 0$ to retrieve the null surface. 
For that purpose we 
start with normal to the $r=\textrm{constant}$ 
surface and then take the null limit. 
With proper choice of  $l_{a}$ and $k_{a}$ 
\cite{Krishna2014} the entropy 
turns out to be:
\begin{equation}\label{Paper2:Eq56}
S=\frac{1}{2}\int d^{D-2}x \sqrt{\mu} P^{ur}_{ur}
\end{equation}
Let us next calculate the surface Hamiltonian that comes from the 
$r=\textrm{constant}$ surface. Surface Hamiltonian has the 
following expression
\begin{equation}\label{Paper2:Eq57}
H_{sur}=\frac{1}{8\pi}\int d^{D-2}x \sqrt{\mu}Q_{a}^{~bdr}\Gamma ^{a}_{bd}
\end{equation}
The only connection components that will remain nonzero  in the $r\rightarrow 0$ limit are those 
given by \eq{Paper2:Eqa09}. With these connections the surface Hamiltonian turns out to be,
\begin{eqnarray}\label{Paper2:Eq58}
H_{sur}&=&\frac{1}{8\pi}\int d^{D-2}x \sqrt{\mu}\Big[2\alpha Q^{ur}_{ur}+2\beta _{A}Q^{Ar}_{ur}+\mu^{BD}\hat{\Gamma}^{A}_{BC}Q_{AD}^{Cr}
-\mu ^{AC}\partial _{r}\mu _{AB}Q^{Br}_{uC}
\nonumber
\\
&+&\frac{1}{2}\left\lbrace -\partial _{u}\mu _{AB}Q_{r}^{~ABr}
+\partial _{u}\mu _{BC}\left(Q^{CBur}+Q^{CuBr}\right)\right\rbrace \Big]
\end{eqnarray}
In the \EH limit only the $Q^{ur}_{ur}$ term contributes leading to \eq{Paper2:Eq40}.
Now using \eq{Paper2:Eq11} all these $Q^{ab}_{cd}$ terms can be calculated. They  lead to:
\begin{eqnarray}
Q^{Ar}_{ur}&=&\delta ^{AruBPQ\ldots}_{urCDRS\ldots}R^{CD}_{uB}R^{RS}_{PQ}\ldots
\label{Paper2:Eq59a}
\\
Q_{AD}^{Cr}&=&\delta ^{CrMNuL\ldots}_{ADurPQ\ldots}R^{ur}_{MN}R^{PQ}_{uL}\ldots 
+\delta ^{CruMPQ\ldots}_{ADurRS\ldots}R^{ur}_{uM}R^{RS}_{PQ}\ldots
\nonumber
\\
&+&\delta ^{CrMPuKUV\ldots}_{ADuQLJrW\ldots}R^{uQ}_{MP}R^{LJ}_{uK}R^{rW}_{UV}\ldots 
+\delta ^{CrMPuLUV\ldots}_{ADuQrKXY\ldots}R^{uQ}_{MP}R^{rK}_{uL}R^{XY}_{UV}\ldots
\nonumber
\\
&+&\delta ^{CruPMNUV\ldots}_{ADuQrLXY\ldots}R^{uQ}_{uP}R^{rL}_{MN}R^{XY}_{UV}\ldots
\label{Paper2:Eq59b}
\\
Q^{Br}_{uC}&=&\delta ^{BruA\ldots}_{uCrD\ldots}R^{rD}_{uA}\ldots 
+\delta ^{BruPJK\ldots}_{uCQRrM\ldots}R^{QR}_{uP}R^{rM}_{JK}\ldots
\label{Paper2:Eq59c}
\\
Q^{Br}_{rA}&=&\delta ^{BruCMN\ldots}_{rAuDPQ\ldots}R^{uD}_{uC}R^{PQ}_{MN}\ldots 
+\delta ^{BruCMN\ldots}_{rAEFuR\ldots}R^{EF}_{uC}R^{uR}_{MN}\ldots 
\label{Paper2:Eq59d}
\\
Q^{CB}_{ru}&=&\delta ^{CBruMN\ldots}_{ruPQRS\ldots}R^{PQ}_{ru}R^{RS}_{MN}\ldots
+\delta ^{CBrPuM\ldots}_{ruQRNS\ldots}R^{MN}_{rP}R^{NS}_{uM}\ldots
\label{Paper2:Eq59e}
\end{eqnarray}
In arriving at the above results we have used the fact that the determinant tensor 
is antisymmetric in any two indices. Thus all the remaining terms, in the above 
expressions, contain only the components of the curvature tensor 
with indices depending on coordinates on the null surface. 
They are all fully characterized by 
the $(D-2)$ metric $\mu _{AB}$. In the null limit we have:
\begin{equation}\label{Paper2:eq60}
R^{CQ}_{uB}=\mu ^{QP}R^{C}_{~PuB};~~ 
R^{rM}_{JK}=\mu ^{MN}R^{r}_{~NJK};~~
R^{rD}_{uC}=-\mu ^{AD}R^{r}_{~ACu};~~
\end{equation}
Having obtained all the components of the surface term we will now consider the thermodynamic 
interpretation.

As we said earlier thermodynamic interpretation can be given provided 
some additional conditions are imposed. As explained in Sec. 
\ref{Paper2:Sec:GNC:EHThermo:Null} we can use 
two type of conditions: either on the variation (with the metric remaining arbitrary) or  
on the metric (keeping the variations arbitrary).
As in the case of \gr, it is preferable to impose the conditions the variations but
keeping the background metric arbitrary. From the result we had in the 
\EH action we would expect these conditions to include
$\delta \left(\mu ^{CA}\partial _{i}\mu _{AB} \right)=0$, 
$\hat{D}_{M}\delta \alpha =0$ 
and $\partial _{i}\delta \mu _{AB}=0$. It turns out that, along with 
these conditions we also need to set $\delta \beta _{A}=0$. 
With these conditions we get the only non-zero contributing term as:
\begin{equation}
H_{sur}=\frac{1}{4\pi}\int d^{D-2}x~\sqrt{\mu}\alpha Q^{ur}_{ur}
=\frac{1}{m}\int d^{D-2}x~\left(\frac{\alpha}{2\pi}\right) 
\left(\frac{1}{2}\sqrt{\sigma} P^{ur}_{ur}\right)=\int d^{D-2}x Ts
\end{equation}
Hence the variation of the surface Hamiltonian leads to:
\begin{equation}\label{Paper2:Eq61}
\delta H_{sur}=-\frac{1}{4\pi}\int d^{D-2}x\left[\delta\left(\sqrt{\mu} Q^{ur}_{ur}\right)\alpha
+\sqrt{\mu} Q^{ur}_{ur}\delta \alpha \right]
\end{equation}
As well as we have the following expression from \eq{Paper2:Eq48},
\begin{equation}\label{Paper2:Eq62}
\delta H_{sur}^{(1)}=\frac{1}{16\pi}\int d^{D-2}x~n_{c}U_{a}^{~bdc}\delta \Gamma ^{a}_{bd}
=\frac{1}{4\pi}\int d^{D-2}x \sqrt{\mu} Q^{ur}_{ur}\delta \alpha
\end{equation}
This along with \eqs{Paper2:Eq61} and (\ref{Paper2:Eq48})  leads to,
\begin{equation}\label{Paper2:Eq63}
\delta H_{sur}^{(2)}=\frac{1}{16\pi}\int d^{D-1}x~n_{c}\Gamma ^{a}_{bd}\delta U_{a}^{~bdc}
=\frac{1}{4\pi}\int d^{D-1}x~\alpha \delta \left(\sqrt{\mu} Q^{ur}_{ur}\right)
\end{equation}
Then \eqs{Paper2:Eq62} and (\ref{Paper2:Eq63}) can also be interpreted in terms of entropy density 
$s=(m/2)\sqrt{\mu} Q^{ur}_{ur}$ leading to:
\begin{eqnarray}
\delta H_{sur}^{(1)}&=&\int d^{D-2}x \frac{\sqrt{\mu} Q^{ur}_{ur}}{2}
\delta \left(\frac{\alpha}{2\pi}\right)
\nonumber
\\
&=&\frac{1}{m}\int d^{D-2}x~s\delta T
\label{Paper2:Eq64a}
\\
\delta H_{sur}^{(2)}&=&
\int d^{D-2}x~\frac{\alpha}{2\pi} \delta \left(\frac{\sqrt{\mu} Q^{ur}_{ur}}{2}\right)
\nonumber
\\
&=&\frac{1}{m}\int d^{D-2}x~T\delta s
\label{Paper2:Eq64b}
\end{eqnarray}
The above results hold for $m$th order \LL Lagrangian, 
which can be generalized in a straight forward manner to a general \LL Lagrangian made of sum of individual \LL Lagrangians of different order.

For the sake of completeness, we mention that
the same results can also be obtained for arbitrary variations if we impose a 
constraint on the background metric. 
This constraint is identical in \gr\ and in \LL models and is given by 
$\partial _{u}\mu _{AB}=0$ and $\hat{D}_{M}\alpha =\partial _{u}(\beta _{M}/2)$. 
This is evident from the components 
of curvature tensor presented in \eqs{Paper2:Eqa10}, (\ref{Paper2:Eqa11}) and (\ref{Paper2:Eqa12}). 
As all of them vanish, when we impose the condition $\partial _{u}\mu _{AB}=0$ condition, from 
\eqs{Paper2:Eq59a}, (\ref{Paper2:Eq59b}), (\ref{Paper2:Eq59c}) 
and \eq{Paper2:eq60} we observe that only $Q^{ur}_{ur}$ term 
remains. This again leads to the standard expressions as in 
\eqs{Paper2:Eq64a} and (\ref{Paper2:Eq64b}).

\section{Describing Gravity in terms of Conjugate Variables}
\label{Paper2:Sec:GravLL}

In the previous section we have shown how variations of the two variables 
introduced in Sec. \ref{Paper2:Sec:ConjuLL:LL} are related to 
variations of temperature and entropy respectively. 
Having addressed the thermodynamic features for these 
variables we now study whether the  
gravitational dynamics can be described in terms of these 
variables.

\subsection{\EH Action with the new set of variables}\label{Paper2:Sec:GravLL:EH}

We start with the  
derivation of equations of motion in terms of 
the variables $f^{ij}$ and $N^{k}_{ij}$ used to describe \EH action 
which was pointed out in \cite{Krishna2013}. 
We shall consider a Palatini-type  variation 
in which  we have taken the two conjugate variables, 
$f^{ij}$ and $N^{k}_{ij}$ as independent. 
For our purpose we shall define the Hamiltonian functional as
\begin{equation}\label{Paper2:Eq67}
\mathcal{H}_{g}=f^{ab}\left(N^{c}_{ad}N^{d}_{bc}-\frac{1}{3}N^{c}_{ac}N^{d}_{bd}\right).
\end{equation}
Then the usual Hamilton's equations of motion obtained from the above Hamiltonian 
will be:
\begin{eqnarray}
\partial _{c}f^{ab}&=&\frac{\partial \mathcal{H}_{g}}{\partial N^{c}_{ab}}
\nonumber
\\
&=&f^{ad}N^{b}_{cd}+f^{bd}N^{a}_{cd}
-\frac{1}{3}f^{am}N^{d}_{dm}\delta ^{b}_{c}-\frac{1}{3}f^{bm}N^{d}_{dm}\delta ^{a}_{c}
\label{Paper2:Eq68a}
\\
\partial _{c}N^{c}_{ab}&=&-\frac{\partial \mathcal{H}_{g}}{\partial f^{ab}}
\nonumber
\\
&=&-N^{c}_{ad}N^{d}_{bc}+\frac{1}{3}N^{c}_{ac}N^{d}_{bd}
\label{Paper2:Eq68b}
\end{eqnarray}
It is easy to show that these equations lead to (i) the correct relation between $f^{ab}$ and 
$N^{c}_{ab}$ and (ii) vacuum Einstein's equations 
$R_{ab}=0$. (For details, see Ref. \cite{Krishna2013}).
(It is this result which actually justifies \textit{a posteriori} our calling $\mathcal{H}_{g}$ the ``Hamiltonian" and \eq{Paper2:Eq68b} the `Hamilton's equations'; the structure is quite different from the corresponding ideas in standard ADM approach; see \cite{Padmanabhan2013a} for more details of this approach.)

We shall now try to generalize these ideas to \LL theories. As we have already noted  in Sec. \ref{Paper2:Sec:ConjuLL:LL}, the natural generalization of these two variables 
does \textit{not} work in \LL models and we need to use the variable introduced in \eq{Paper2:Eq26b}. It is therefore important to first verify what happens to \gr\ when we use these variables.
So we will first study the \EH case before proceeding to \LL theory.
For this purpose we shall start with the \EH Lagrangian written as:
\begin{eqnarray}\label{Paper2:Eq73}
L&=&\sqrt{-g}R=\sqrt{-g}\frac{1}{2}\left(g^{bd}\delta _{a}^{c}-g^{bc}\delta _{a}^{d}\right)R^{a}_{~bcd}
\nonumber
\\
&=&\sqrt{-g}\left(g^{bd}\delta _{a}^{c}-g^{bc}\delta _{a}^{d}\right)
\left(\partial _{c}\Gamma ^{a}_{bd}-\Gamma ^{a}_{dp}\Gamma ^{p}_{bc}\right)
\nonumber
\\
&\equiv &U_{a}^{~bcd}\left(\partial _{c}\Gamma ^{a}_{bd}-\Gamma ^{a}_{dp}\Gamma ^{p}_{bc}\right)
\end{eqnarray}
where we have defined the variable, 
$U_{a}^{~bcd}=\sqrt{-g}\left(g^{bd}\delta _{a}^{c}-g^{bc}\delta _{a}^{d}\right)$ 
having all the symmetries of curvature tensor. Let us see what happens if we treat 
the variables $U_{a}^{~bcd}$ and 
$\Gamma ^{a}_{bc}$ as independent as befitting conjugate variables. The variation of $\Gamma ^{a}_{bc}$ leads to:
\begin{eqnarray}\label{Paper2:Eq74}
\delta L\mid _{U_{a}^{~bcd}}&=&
U_{a}^{~bcd}\left(\partial _{c}\delta \Gamma ^{a}_{bd}-\Gamma ^{p}_{bc}\delta \Gamma ^{a}_{dp}
-\Gamma ^{a}_{dp}\delta \Gamma ^{p}_{bc}\right)
\nonumber
\\
&=&U_{a}^{~bcd}\nabla _{c}\delta \Gamma ^{a}_{bd}
\end{eqnarray}
In order to get to the final expression 
we have used the fact that the Lagrangian is a scalar density and hence can be evaluated in 
a local inertial frame with $U_{a}^{~bcd}$ and $\delta \Gamma ^{a}_{bc}$ as tensors. (One can explicitly verify this result, even without using this trick.)
Then we can rewrite \eq{Paper2:Eq74} as:
\begin{eqnarray}\label{Paper2:Eq75}
\delta L\mid _{U_{a}^{~bcd}}
&=&\nabla _{c}\left(U_{a}^{~bcd}\delta \Gamma ^{a}_{bd}\right)
-\delta \Gamma ^{a}_{bd}\nabla _{c}U_{a}^{~bcd}
\end{eqnarray}
When the above equation integrated over the spacetime volume to obtain the variation $\delta L$ of the action, the first term 
$\nabla _{c}\left(U_{a}^{~bcd}\delta \Gamma ^{a}_{bd}\right)$, being a total 
divergence contributes only on the surface and hence can be dropped. Thus the condition $\delta L=0$ for arbitrary variations $\delta \Gamma ^{a}_{bc}$ leads to:
\begin{equation}\label{Paper2:Eq76}
\nabla _{c}U_{a}^{~bcd}=0
\end{equation}
From the definition of $U_{a}^{~bcd}$ it is evident that \eq{Paper2:Eq76} 
implies $\nabla _{c}g_{ab}=0$, standard result from Palatini variation in 
\EH action \cite{Padmanabhan2010b} [This result can also be obtained using \eq{Paper2:Eqanew01}].

But we are led to a difficulty in this approach when we vary $U_{a}^{~bcd}$ 
because the variation of $U_{a}^{~bcd}$ leads 
to $R^{a}_{~bcd}=0$, i.e. flat spacetime! It is not possible to get Einstein equation 
from \textit{arbitrary} variations of $U_{a}^{~bcd}$ since it has four indices, naturally 
leading to zero curvature tensor, equivalently flat spacetime. 
The reason for this disaster is simple. When we vary $U_{a}^{~bcd}$ we are pretending that we are varying 20 independent components (because $U_{a}^{~bcd}$ has the symmetries of curvature tensor); but we know that --- since $U_{a}^{~bcd}$ is completely determined by $g_{ab}$ --- it really has only  10 independent components.
So, in order to get correct equations of motion 
we need to restrict variations such that there are only  10 of them 
independent  components in the variation $\delta U_{a}^{~bcd}$. This is easy to achieve. Since  an arbitrary symmetric 
second rank tensor $S^{p}_{q}$ has 10 independent components, we can easily construct such a constrained variation by considering a subclass of$\delta U_{a}^{~bcd}$ which is determined by the variations $\delta S^{p}_{q}$ of an arbitrary second rank tensor. This leads us to consider 
variations of the form:
\begin{equation}\label{Paper2:Eq77}
\delta U_{p}^{~qrs}=\left(U_{m}^{~qrs}\delta ^{n}_{p}
-\frac{1}{2}U_{p}^{~qrs}\delta ^{n}_{m}\right)\delta S^{m}_{n}
\end{equation}
(In fact, it turns out that the above variation can be slightly generalized by 
introducing a sixth rank tensor $A_{pm}^{~~qrsn}$ satisfying the criteria 
$A_{pm}^{~~qrsn}R^{p}_{~qrs}=0$. We shall not consider 
these variations any more since they have no effect on the 
equation of motion.) With these restricted class of variations we arrive at:
\begin{eqnarray}\label{Paper2:Eq78}
\delta L\mid _{\Gamma ^{a}_{bc}}&=&\frac{1}{2}R^{a}_{~bcd}\delta U_{a}^{~bcd}
\nonumber
\\
&=&\frac{1}{2}R^{p}_{~qrs}\left(U_{m}^{~qrs}\delta ^{n}_{p}
-\frac{1}{2}\delta ^{n}_{m}U_{p}^{~qrs}\right)\delta S^{m}_{n}
\end{eqnarray}
Then for arbitrary variations of the symmetric tensor 
$S^{a}_{b}$, we get the equations of motion:
\begin{equation}\label{Paper2:Eq79}
R^{p}_{~qrs}\left(U_{m}^{~qrs}\delta ^{n}_{p}
-\frac{1}{2}\delta ^{n}_{m}U_{p}^{~qrs}\right)=0
\end{equation}
To prove the equivalence with Einstein 
equation we note that the following relations
\begin{equation}\label{Paper2:Eq80}
R^{p}_{~qrs}U_{m}^{~qrs}=2\sqrt{-g}R^{p}_{m};~~~R^{p}_{~qrs}U_{p}^{~qrs}=2\sqrt{-g}R
\end{equation}
directly transform the equation of motion (\ref{Paper2:Eq79})  to, $G_{ab}=0$, 
the source-free Einstein equation. (We have not included matter fields to our 
system about which we shall comment later.)
In the next section we will show the validity of the above formalism for \LL gravity. 

\subsection{Generalization to \LL gravity}
\label{Paper2:Sec:GravLL:LL}

In the case of \LL\ models the appropriate Lagrangian to consider 
for our purpose is:
\begin{equation}\label{Paper2:Eq81}
L=U_{a}^{~bcd}\left(\partial _{c}\Gamma ^{a}_{bd}-\Gamma ^{a}_{md}\Gamma ^{m}_{bc}\right)
\end{equation}
which, using \eq{Paper2:Eq26b}, can be identified with the \LL Lagrangian. 
The variation of the above Lagrangian 
with respect to $\Gamma ^{a}_{bc}$ leads to
\begin{equation}\label{Paper2:Eq82}
\nabla _{c}U_{a}^{~bcd}=0
\end{equation}
as in the \EH scenario (see \eq{Paper2:Eq76}). 
This condition is equivalent to the criteria 
that in \LL gravity $\nabla _{c}P^{abcd}=0$. \textit{This result 
is quiet remarkable}, since the criterion that the field equation should 
be of second order in the dynamical variable  gets 
into picture automatically from variation 
of the Lagrangian. In fact, this condition has another aspect to it.
In general, when we study the metric formulation we treat the  
Lagrangian with $g_{ab}$ as the independent variable (with connections given in terms of the metric) while in the Palatini formulation,  we treat both the 
metric and the connections as independent and their variation leads to the 
relation between them \textit{and} the equation of motion. For an
arbitrary Lagrangian the metric
and Palatini variation do not coincide \cite{Exirifard2008}. However 
if the condition $\nabla _{c}\left(\partial L/\partial R_{abcd}\right)=0$ is satisfied 
then both the metric and Palatini formulations coincide. This is identical 
to the condition presented in \eq{Paper2:Eq09} and it is interesting to see this condition emerging from a variation here.

Next we need to vary the Lagrangian with respect to  $U_{a}^{~bcd}$. Arbitrary variation of $U_{a}^{~bcd}$ treating all the 20 components independent will 
lead to trouble, just as in \gr. Since the Lagrangian in \eq{Paper2:Eq81} can equivalently be written as 
$L=(1/2) U_{a}^{~bcd}R^{a}_{~bcd}$, such that for arbitrary variation of $U_{a}^{~bcd}$ we get 
$R^{a}_{~bcd}=0$, i.e. flat spacetime solution --- just as in \gr. 
In order to get the field equation we need to again consider only a subclass of variations 
as we did in the \EH scenario to derive the equations of motion in \eq{Paper2:Eq79}. 
Here again 
we need to assume that not all the independent components of $U_{a}^{~bcd}$ are contributing 
to the variation but only 10 degrees of freedom, which can be encoded by a symmetric second rank part with arbitrary variation. This 
amounts to taking:
\begin{equation}\label{paper2:Eq83}
\delta U_{p}^{~qrs}=\left(U_{m}^{~qrs}\delta ^{n}_{p}
-\frac{1}{2}U_{p}^{~qrs}\delta ^{n}_{m}\right)\delta S^{m}_{n}
\end{equation}
(Here also we can introduce an additional sixth rank tensor as we 
did after \eq{Paper2:Eq77}. However as far as the equation of motion 
is concerned it has no effect and thus will not be considered 
any more.) With these restricted class of variations 
the Lagrangian variation leads to:
\begin{equation}\label{Paper2:Eq84}
\delta L\mid _{\Gamma ^{a}_{bc}}=\left(mU_{a}^{~pqr}\delta ^{b}_{s}
-\frac{1}{2}\delta ^{b}_{a}U_{s}^{~pqr}\right)R^{s}_{~pqr}\delta S^{a}_{b}
\end{equation}
where $\delta S^{a}_{b}$ is variation of an arbitrary symmetric second rank tensor and 
the factor $m$ comes from the fact that we are considering $m$th order \LL Lagrangian.
When the variation $\delta S^{a}_{b}$ is considered arbitrary the equation of motion turns out to be
\begin{equation}\label{Paper2:Eq85}
\left(mU_{a}^{~pqr}\delta ^{b}_{s}
-\frac{1}{2}\delta ^{b}_{a}U_{s}^{~pqr}\right)R^{s}_{~pqr}=0.
\end{equation}
To show that the above equation of motion is indeed identical to the equation of motion 
in \LL gravity we just use \eq{Paper2:Eq26b} to substitute for $U_{a}^{~bcd}$ 
leading to:
\begin{eqnarray}
0&=&mQ_{a}^{~pqr}R^{b}_{~pqr}-\frac{1}{2}\delta ^{b}_{a}Q_{s}^{~pqr}R^{s}_{~pqr}
\nonumber
\\
&=&\mathcal{R}^{a}_{b}-\frac{1}{2}\delta ^{a}_{b}L
\end{eqnarray}\label{Paper2:Eq86}
which is the \LL equation of motion. Thus we observe that these 
two variables satisfy all the criteria that conjugate variables should.

The above result is derived for $m$th order \LL Lagrangian and can be easily 
generalized to general \LL Lagrangian $L=\sum _{m}c_{m}L^{(m)}$. Then the above 
variation of \LL Lagrangian leads to the following 
expression:
\begin{eqnarray}\label{Paper2:Eq87}
\delta \left(\sqrt{-g}L\right)|_{\Gamma ^{a}_{bc}}&=&\sum _{m}c_{m}
\delta \left(\sqrt{-g} L^{(m)}\right)|_{\Gamma ^{a}_{bc}}
\nonumber
\\
&=&\left\lbrace \left(\sum _{m}c_{m} mU_{a}^{~pqr}\right)\delta ^{b}_{s}
-\frac{1}{2}\delta ^{b}_{a}\left(\sum _{m}c_{m}U_{s}^{~pqr}\right)\right\rbrace 
R^{s}_{~pqr}\delta S^{a}_{b}
\end{eqnarray}
For arbitrary variation of the symmetric tensor $S^{a}_{b}$ the equation 
of motion can be obtained as:
\begin{eqnarray}\label{Paper2:Eq88}
\left\lbrace \left(\sum _{m}c_{m} mU_{a}^{~pqr}\right)\delta ^{b}_{s}
-\frac{1}{2}\delta ^{b}_{a}\left(\sum _{m}c_{m}U_{s}^{~pqr}\right)\right\rbrace 
R^{s}_{~pqr}=0
\end{eqnarray}
Note that with the following relations
\begin{eqnarray}
\sum _{m}c_{m} mU_{a}^{~pqr}=\frac{\partial \sqrt{-g} L}{\partial R^{a}_{~pqr}}
=\sqrt{-g}P_{a}^{~pqr}
\label{Paper2:Eq89a}
\\
\sum _{m}c_{m}U_{s}^{~pqr}R^{s}_{~pqr}=\sqrt{-g}L
\label{Paper2:Eq89b}
\end{eqnarray}
the above \eq{Paper2:Eq88} becomes equivalent to
\begin{eqnarray}\label{Paper2:Eq90}
\mathcal{R}^{b}_{a}-\frac{1}{2}\delta ^{b}_{a}L=0
\end{eqnarray}
which is the \LL field equation. 

In this approach, we arrive at the 
condition that needs to be imposed in order to get second order equation of motion, directly from a variational principle along with 
the field equation. The price we pay is the following: (a) We need to restrict the form of the variations, the physical meaning of which is unclear. (b) The inclusion of matter 
in this scheme is difficult. Usually, the energy momentum tensor comes  from the variation of the 
matter Lagrangian with respect to the metric alone and since we have not included the metric 
in our formulation it is not clear how to include matter. These  issues require further 
investigation. 

\section{Concluding Remarks}\label{Paper2:Sec:Conc}

The link between the standard approach to gravity and the thermodynamical one is provided by the action principle of gravity. Previous studies have shown that these actions have several peculiar features and --- under suitable conditions --- a thermodynamical interpretation. This motivates us to look for geometrical variables in which the expression for action simplifies and which will have direct thermodynamical interpretation. More specifically, we want to discover geometrical variables, symbolically called $[q,p]$ such that $q\delta p$ and $p\delta q$ will correspond to $s\delta T$ and $T\delta s$ where $T$ is the horizon temperature and $s$ is the entropy density. 

This goal was achieved for the \EH\ action recently \cite{Krishna2013}
 by introducing canonically 
conjugate variables as $f^{ab}=\sqrt{-g}g^{ab}$ and the corresponding 
momenta $N^{c}_{ab}$. In terms of these 
variables, the surface term turns out to have the structure $-\partial (qp)$ and the their variations have direct thermodynamic interpretation.

It has been noticed in the past that virtually every result involving the thermodynamical interpretation gravity, which was valid for \gr\ , could be generalized to \LL\ models. We have shown that this fact holds for the above result as well. We could introduce two suitable variables in the case of \LL models with the following properties: (a) These variables reduce to the ones used in \gr\ in $D=4$ when the \LL model reduces to \gr. (b) The variation of these quantities correspond to $s\delta T$ and $T\delta s$ where $s$ is now the correct Wald entropy density of the \LL model. This 
result holds rather trivially on any static (but not necessarily spherically symmetric or matter-free) horizon and --- more importantly --- on any 
 arbitrary null surface acting 
as local Rindler horizon. Since local Rindler structures can be imposed on any event, this shows that, around any event, certain geometric variables can be attributed thermodynamical significance.

These variables, by themselves, seem interesting and deserves further study. For example, we found that they can be thought of as connections and conjugate momenta associated with connections in a formal sense. But to get  sensible equations of motion by varying these quantities, we needed to restrict their variation in a manner which --- while mathematically rigorous --- is physically unclear. We thus find that while the \textit{thermodynamic} significance of these variables are clear and direct, the \textit{dynamical} significance requires further work to establish.

The analysis once again confirms that the thermodynamic interpretation goes far deeper than \gr\ and is definitely telling us something nontrivial about the structure of the spacetime. We note that the nature of Wald entropy density in \LL\ models is far more complicated than a simple constant (1/4) in \gr\ ; yet, everything works out exactly as expected.
The action principle 
somehow encodes the information about horizon thermodynamics, which is a key 
result in emergent gravity paradigm.

\section*{Acknowledgement}

Research of T.P is partially supported by J.C. Bose research grant of DST, Govt. of India. 
Research of S.C 
is funded by a SPM Fellowship from CSIR, Govt. of India. 
S.C also likes to thank Krishnamohan Parattu, Suprit Singh, Bibhas Ranjan Majhi 
and Kinjalk Lochan
for helpful discussions. We thank Naresh Dadhich for useful comments.

\appendix

\section{Derivation of Various Identities used in Text}\label{Paper2:AppA}

We will consider the $p\partial q$ and $q\partial p$ structure arising from the identification of 
$\tilde{f}^{ab}$ as coordinate and $\tilde{N}^{c}_{ab}$ as momentum in \LL gravity. For 
the calculation, the following identity will be used here and there:
\begin{eqnarray}\label{Paper2:Eqa01}
0=\nabla _{c}Q_{ab}^{cd}&=&\partial _{c}Q_{ab}^{cd}+\Gamma ^{c}_{ck}Q_{ab}^{kd}
\nonumber
\\
&-&\Gamma ^{k}_{ca}Q_{kb}^{cd}-\Gamma ^{k}_{cb}Q_{ak}^{cd}+\Gamma ^{d}_{ck}Q_{ab}^{ck}
\end{eqnarray}
However $Q_{ab}^{cd}$ being antisymmetric in (c,d) while $\Gamma ^{c}_{ab}$ being symmetric 
in (a,b) the last term in the above expansion vanishes. Thus ordinary derivative of the quantity 
$Q_{ab}^{cd}$ has the following expression
\begin{equation}\label{Paper2:Eqa02}
\partial _{c}Q_{ab}^{cd}=-\Gamma ^{c}_{ck}Q_{ab}^{kd}+\Gamma ^{k}_{ca}Q_{kb}^{cd}
+\Gamma ^{k}_{cb}Q_{ak}^{cd}
\end{equation}
Note that we can include $\sqrt{-g}$ in the above expression leading to:
\begin{equation}\label{Paper2:Eqanew01}
\partial _{c}\left(\sqrt{-g}Q_{a}^{~bcd}\right)=\left(\sqrt{-g}Q_{p}^{~bcd}\right)\Gamma ^{p}_{ac}-
\left(\sqrt{-g}Q_{a}^{pcd}\right)\Gamma ^{b}_{cp}
\end{equation}
Thus we get the following expression from \eq{Paper2:Eqa02}:
\begin{eqnarray}\label{Paper2:Eqa03}
\partial _{c}\tilde{N}^{c}_{ab}&=&\partial _{c}
\left[Q_{bp}^{cq}\Gamma ^{p}_{aq}+Q_{ap}^{cq}\Gamma ^{p}_{bq} \right]
\nonumber
\\
&=& \left(\partial _{c}Q_{bp}^{cq} \right)\Gamma ^{p}_{aq}+Q_{bp}^{cq}\partial _{c}\Gamma ^{p}_{aq}
+\left(\partial _{c}Q_{ap}^{cq}\right)\Gamma ^{p}_{bq}+Q_{ap}^{cq}\partial _{c}\Gamma ^{p}_{bq}
\nonumber
\\
&=&Q_{kp}^{cq}\Gamma ^{k}_{cb}\Gamma ^{p}_{aq}+Q_{bk}^{cq}\Gamma ^{k}_{cp}\Gamma ^{p}_{aq}
-Q_{bp}^{kq}\Gamma ^{p}_{aq}\Gamma ^{c}_{ck}+Q_{kp}^{cq}\Gamma ^{k}_{ca}\Gamma ^{p}_{bq}
\nonumber
\\
&+&Q_{ak}^{cq}\Gamma ^{k}_{cp}\Gamma ^{p}_{bq}-Q_{ap}^{kq}\Gamma ^{c}_{ck}\Gamma ^{p}_{bq}
+Q_{bp}^{cq}\partial _{c}\Gamma ^{p}_{aq}+Q_{ap}^{cq}\partial _{c}\Gamma ^{p}_{bq}
\end{eqnarray}
where in order to arrive at the last equality \eq{Paper2:Eqa02} has been used. Now contracting 
the above expression with $\tilde{f}^{ab}$ we readily obtain
\begin{eqnarray}\label{Paper2:Eqa04}
\tilde{f}^{ab}\partial _{c}\tilde{N}^{c}_{ab}&=&\sqrt{-g}g^{ab}
\left[2Q_{ap}^{cq}\partial _{c}\Gamma ^{p}_{bq}+2Q_{kp}^{cq}\Gamma ^{k}_{ca}\Gamma ^{p}_{bq}
+2Q_{ak}^{cq}\Gamma ^{k}_{cp}\Gamma ^{p}_{bq}-2Q_{ap}^{kq}\Gamma ^{c}_{ck}\Gamma ^{p}_{bq} \right]
\nonumber
\\
&=&-2\sqrt{-g}Q_{p}^{~bcq}\left(\partial _{c}\Gamma ^{p}_{bq}+\Gamma ^{p}_{ck}\Gamma ^{k}_{bq} \right)
\nonumber
\\
&+&2\sqrt{-g}Q_{p}^{~bkq}\Gamma ^{c}_{ck}\Gamma ^{p}_{bq}
+2\sqrt{-g}g^{ab}Q_{kp}^{cq}\Gamma ^{k}_{ca}\Gamma ^{p}_{bq}
\nonumber
\\
&=&-\sqrt{-g}Q_{p}^{~bqc}R^{p}_{~bqc}+2\sqrt{-g}Q_{p}^{~bkq}\Gamma ^{c}_{ck}\Gamma ^{p}_{bq}
+2\sqrt{-g}g^{ab}Q_{kp}^{cq}\Gamma ^{k}_{ca}\Gamma ^{p}_{bq}.
\end{eqnarray}
Note that in the \EH limit the last two terms adds up to yield $-\sqrt{-g}L_{quad}$. 
Then consider the other 
combination which can be expressed as
\begin{eqnarray}\label{Paper2:Eqa05}
\tilde{N}^{c}_{ab}\partial _{c}\tilde{f}^{ab}&=&\left(Q_{ap}^{cq}\Gamma ^{p}_{qb}
+Q_{bp}^{cq}\Gamma ^{p}_{qa}\right)\partial _{c}\left(\sqrt{-g}g^{ab}\right)
\nonumber
\\
&=&\sqrt{-g}\left(Q_{ap}^{cq}\Gamma ^{p}_{qb}+Q_{bp}^{cq}\Gamma ^{p}_{qa}\right)
\left(\partial _{c}g^{ab}+g^{ab}\Gamma ^{p}_{cp} \right)
\nonumber
\\
&=&2\sqrt{-g}Q_{ap}^{cq}\Gamma ^{p}_{qb}\partial _{c}g^{ab}
+2\sqrt{-g}Q_{p}^{~bqc}\Gamma ^{p}_{qb}\Gamma ^{m}_{cm}
\nonumber
\\
&=&2\sqrt{-g}Q_{p}^{~bcq}\Gamma ^{l}_{bc}\Gamma ^{p}_{ql}
+2\sqrt{-g}Q_{p}^{~bqc}\Gamma ^{p}_{qb}\Gamma ^{m}_{cm}
-2\sqrt{-g}g^{bm}Q_{ap}^{cq}\Gamma ^{p}_{qb}\Gamma ^{a}_{cm}
\nonumber
\\
&=&\sqrt{-g}L_{quad}+2\sqrt{-g}Q_{p}^{~bqc}\Gamma ^{p}_{qb}\Gamma ^{m}_{cm}
-2\sqrt{-g}g^{bm}Q_{ap}^{cq}\Gamma ^{p}_{qb}\Gamma ^{a}_{cm}.
\end{eqnarray}
In the \EH limit the above term leads to $2\sqrt{-g}L_{quad}$. 
Next we will derive similar relations which actually 
behaves as conjugate variables, with the identification, 
$p\equiv 2\sqrt{-g}Q_{a}^{~bcd}$ and $q\equiv \Gamma^{a}_{bc}$. Then the respective $p\partial q$ 
and $q\partial p$ expressions are given in the following results:
\begin{eqnarray}\label{Paper2:Eqa06}
2\sqrt{-g}Q_{e}^{~bdc}\partial _{c}\Gamma ^{e}_{bd}&=&\sqrt{-g}Q_{e}^{~bdc}
\left(\partial _{c}\Gamma ^{e}_{bd}-\partial _{d}\Gamma ^{e}_{bc} \right)
\nonumber
\\
&=&\sqrt{-g}Q_{e}^{~bdc}R^{e}_{~bcd}-2\sqrt{-g}Q_{e}^{~bdc}\Gamma ^{e}_{mc}\Gamma ^{m}_{bd}
\nonumber
\\
&=&-\sqrt{-g}Q_{e}^{~abc}R^{e}_{~abc}-\sqrt{-g}L_{quad}
\end{eqnarray}
and
\begin{eqnarray}\label{Paper2:Eqa07}
\Gamma ^{d}_{be}\partial _{c}\left(2\sqrt{-g}Q_{d}^{~bec} \right)&=&
2\sqrt{-g}\Gamma ^{d}_{be}\partial _{c}Q_{d}^{~bec}+2\Gamma ^{d}_{be}Q_{d}^{~bec}\partial _{c}\sqrt{-g}
\nonumber
\\
&=& 2\sqrt{-g}\Gamma ^{d}_{be}\left(\Gamma ^{a}_{cd}Q_{a}^{~bec}-\Gamma ^{b}_{ca}Q_{d}^{~aec}
-\Gamma ^{c}_{ca}Q_{d}^{~bea} \right)+2\Gamma ^{d}_{be}Q_{d}^{~bec}\partial _{c}\sqrt{-g}
\nonumber
\\
&=&2\sqrt{-g}L_{quad}
\end{eqnarray}
Now we will show one derivative used in the text for \LL Lagrangian:
\begin{eqnarray}\label{Paper2:Eqanew}
\frac{\partial \left(\sqrt{-g}L\right)}{\partial \left(\partial _{l}\Gamma ^{u}_{vw}\right)}
&=&m\sqrt{-g}\delta ^{aba_{2}b_{2}\ldots a_{m}b_{m}}_{cdc_{2}d_{2}\ldots c_{m}d_{m}}
\frac{\partial R^{cd}_{ab}}{\partial \left(\partial _{l}\Gamma ^{u}_{vw}\right)}
R^{c_{2}d_{2}}_{a_{2}b_{2}}\ldots R_{a_{m}b_{m}}^{c_{m}d_{m}}
\nonumber
\\
&=&m\sqrt{-g}\delta ^{aba_{2}b_{2}\ldots a_{m}b_{m}}_{cdc_{2}d_{2}\ldots c_{m}d_{m}}
\left[ g^{dp}\delta ^{c}_{u}\delta ^{v}_{p}\left(\delta ^{l}_{a}\delta ^{w}_{b}-
\delta ^{l}_{b}\delta ^{w}_{a}\right)\right]R^{c_{2}d_{2}}_{a_{2}b_{2}}\ldots R_{a_{m}b_{m}}^{c_{m}d_{m}}
\nonumber
\\
&=&2m\sqrt{-g}g^{dv}Q^{lw}_{ud}=mU_{u}^{~vlw}
\end{eqnarray}
The next thing to consider are the connections in the general static metric we are considering. 
There the connections are given by
\begin{eqnarray}\label{Paper2:Eqa08}
\Gamma ^{n}_{nn}&=&\Gamma ^{n}_{tn}=\Gamma ^{n}_{nA}=\Gamma ^{n}_{tA}=0
\nonumber
\\
\Gamma ^{n}_{tt}&=&N\partial _{n}N;~~\Gamma ^{n}_{AB}=-\frac{1}{2}\partial _{n}\sigma _{AB}
\nonumber
\\
\Gamma ^{t}_{nn}&=&\Gamma ^{t}_{nA}=0
\nonumber
\\
\Gamma ^{t}_{nt}&=&\frac{\partial _{n}N}{N};~~\Gamma ^{t}_{tt}=\frac{\partial _{t}N}{N}
\nonumber
\\
\Gamma ^{t}_{tA}&=&\frac{\partial _{A}N}{N};~~\Gamma ^{t}_{AB}=\frac{\partial _{t}\sigma _{AB}}{2N^{2}}
\nonumber
\\
\Gamma ^{A}_{nn}&=&\Gamma ^{A}_{nt}=0
\nonumber
\\
\Gamma ^{A}_{nB}&=&\frac{1}{2}\sigma ^{AC}\left(\partial _{n}\sigma _{BC}\right);~~\Gamma ^{A}_{tt}=N\sigma ^{AB}\partial _{B}N
\nonumber
\\
\Gamma ^{A}_{tB}&=&\frac{1}{2}\sigma ^{AC}\left(\partial _{t}\sigma _{BC}\right);~~
\Gamma ^{A}_{BC}=\frac{1}{2}\sigma ^{AD}\left(-\partial _{D}\sigma _{BC}+\partial _{B}\sigma _{CD}+\partial _{C}\sigma _{BD}\right)
\end{eqnarray}
Also we list below all the connections in GNC coordinate system that will remain nonzero in the 
null surface limit:
\begin{eqnarray}\label{Paper2:Eqa09}
\Gamma ^{u}_{uu}&=&\alpha;~~\Gamma ^{u}_{uA}=\beta _{A}/2;~~\Gamma ^{u}_{AB}=-\partial _{r}\mu _{AB}/2
\nonumber
\\
\Gamma ^{r}_{ur}&=&-\alpha;~~\Gamma ^{r}_{rA}=-\beta _{A}/2;~~\Gamma ^{r}_{AB}=-\partial _{u}\mu _{AB}/2
\nonumber
\\
\Gamma ^{A}_{BC}&=&\hat{\Gamma}^{A}_{BC};~~\Gamma ^{A}_{Bu}=\mu ^{CA}\partial _{u}\mu _{BC}/2;~~ 
\nonumber
\\
\Gamma ^{A}_{Br}&=&\mu ^{CA}\partial _{r}\mu _{BC}/2;~~\Gamma ^{A}_{ur}=-\beta ^{A}/2
\end{eqnarray}
Now we will present curvature tensor components in GNC coordinates, which are relevant 
for the calculations in the main text
\begin{eqnarray}\label{Paper2:Eqa10}
R^{C}_{~PuB}&=&\partial _{u}\Gamma ^{C}_{PB}-\partial _{B}\Gamma ^{C}_{Pu}
+\Gamma ^{C}_{mu}\Gamma ^{m}_{PB}-\Gamma ^{C}_{mB}\Gamma ^{m}_{Pu}
\nonumber
\\
&=&\partial _{u}\left(\frac{1}{2}\bar{\beta}^{C}\partial _{r}\mu _{PB}+\hat{\Gamma}^{C}_{PB} \right)
-\partial _{B}\left(\frac{1}{2}\bar{\beta}^{C}\partial _{r}\bar{\beta}_{P}
+\frac{1}{2}\mu ^{CA}\partial _{u}\mu _{PA}
+\frac{1}{2}\mu ^{CA}\left(\hat{D}_{P}\bar{\beta}_{A}-\hat{D}_{A}\bar{\beta} _{P}\right) \right)
\nonumber
\\
&+&\left(\frac{1}{2}\bar{\beta}^{C}\partial _{r}\bar{\beta}_{A}
+\frac{1}{2}\mu ^{CB}\partial _{u}\mu _{BA}
+\frac{1}{2}\mu ^{CB}\left(\hat{D}_{A}\bar{\beta}_{B}-\hat{D}_{B}\bar{\beta} _{A}\right) \right)
\left(\frac{1}{2}\bar{\beta}^{A}\partial _{r}\mu _{BP}+\hat{\Gamma}^{A}_{BP} \right)
\nonumber
\\
&-&\left(\frac{1}{2}\bar{\beta}^{C}\partial _{r}\bar{\alpha}-\frac{1}{2}\mu ^{CA}\hat{D}_{A}\bar{\alpha}
+\mu ^{AC}\partial _{u}\bar{\beta}_{A}\right)
\left(\frac{1}{2}\partial _{r}\mu _{AB}\right)
\nonumber
\\
&-&\left(\frac{1}{2}\mu ^{CA}\partial _{r}\bar{\beta}_{A}\right)
\left(\frac{1}{2}\left\lbrace \partial _{u}\mu _{PB}
+\left(\bar{\beta}^{C}\bar{\beta}_{C}-\bar{\alpha} \right)\partial _{r}\mu _{PB}\right\rbrace 
+\frac{1}{2}\left(\hat{D}_{P}\bar{\beta} _{B}+\hat{D}_{B}\bar{\beta} _{P} \right) \right)
\nonumber
\\
&-&\left(\frac{1}{2}\bar{\beta}^{C}\partial _{r}\mu _{BA}+\hat{\Gamma}^{C}_{AB} \right)
\left(\frac{1}{2}\bar{\beta}^{A}\partial _{r}\bar{\beta}_{P}+\frac{1}{2}\mu ^{CA}\partial _{u}\mu _{PC}
+\frac{1}{2}\mu ^{CA}\left(\hat{D}_{P}\bar{\beta}_{C}-\hat{D}_{C}\bar{\beta} _{P}\right) \right)
\nonumber
\\
&+&\left(\frac{1}{2}\bar{\beta}^{C}\partial _{r}\bar{\beta}_{B}
+\frac{1}{2}\mu ^{CA}\partial _{u}\mu _{BA}
+\frac{1}{2}\mu ^{CA}\left(\hat{D}_{B}\bar{\beta}_{A}-\hat{D}_{A}\bar{\beta} _{B}\right) \right)
\left(\frac{1}{2}\partial _{r}\bar{\beta} _{P} \right)
\nonumber
\\
&+&\left(\frac{1}{2}\mu ^{CA}\partial _{r}\mu _{BA} \right)
\left(\frac{1}{2}\left(\bar{\beta}^{C}\bar{\beta}_{C}-\bar{\alpha} \right)
\partial _{r}\bar{\beta}_{P}+\frac{1}{2}\hat{D}_{P}\bar{\alpha}
-\frac{1}{2}\bar{\beta}^{B}\left(\partial _{u}\mu _{PB}
+\hat{D}_{P}\bar{\beta}_{B}-\hat{D}_{B}\bar{\beta} _{P} \right) \right)
\nonumber
\\
&=&\partial _{u}\hat{\Gamma}^{C}_{PB}-\frac{1}{2}\mu ^{CM}\partial _{B}\partial _{u}\mu _{MP}
-\frac{1}{2}\partial _{B}\mu ^{CM}\partial _{u}\mu _{MP}
+\frac{1}{2}\mu ^{CM}\partial _{u}\mu _{MA}\hat{\Gamma}^{A}_{PB}
\nonumber
\\
&-&\frac{1}{2}\partial _{u}\mu _{PB}\Gamma ^{C}_{ru}
-\frac{1}{2}\mu ^{CM}\partial _{u}\mu _{MB}\Gamma ^{u}_{Pu}
-\frac{1}{2}\mu ^{AC}\partial _{u}\mu _{CP}\hat{\Gamma}^{C}_{AB}
\end{eqnarray}
\begin{eqnarray}\label{Paper2:Eqa11}
R^{r}_{~ABu}&=&\partial _{B}\Gamma ^{r}_{Au}-\partial _{u}\Gamma ^{r}_{AB}+\Gamma ^{r}_{Bm}\Gamma ^{m}_{Au}-\Gamma ^{r}_{um}\Gamma ^{m}_{BA}
\nonumber
\\
&=&\partial_{B}\left[-\frac{1}{2}\left(\bar{\beta}^{C}\bar{\beta}_{C}-\bar{\alpha} \right)
\partial _{r}\bar{\beta}_{A}+\frac{1}{2}\hat{D}_{A}\bar{\alpha}-\frac{1}{2}\bar{\beta}^{B}
\left(\partial _{u}\mu _{AB}+\hat{D}_{A}\bar{\beta}_{B}-\hat{D}_{B}\bar{\beta} _{A} \right) \right]
\nonumber
\\
&-&\partial _{u}\left[-\frac{1}{2}\left\lbrace \partial _{u}\mu _{AB}+
\left(\bar{\beta}^{C}\bar{\beta}_{C}-\bar{\alpha} \right)\partial _{r}\mu _{AB}\right\rbrace
+\frac{1}{2}\left(\hat{D}_{A}\bar{\beta} _{B}+\hat{D}_{B}\bar{\beta} _{A} \right) \right]
\nonumber
\\
&+&\Big(\left[\frac{1}{2}\left(\partial _{r}\bar{\beta}_{B}-\bar{\beta}^{C}\partial _{r}\mu _{CB} \right) \right]
\left[-\frac{1}{2}\left(\bar{\beta}^{C}\bar{\beta}_{C}-\bar{\alpha} \right)
\partial _{r}\bar{\beta}_{A}+\frac{1}{2}\hat{D}_{A}\bar{\alpha}-\frac{1}{2}\bar{\beta}^{B}
\left(\partial _{u}\mu _{AB}+\hat{D}_{A}\bar{\beta}_{B}-\hat{D}_{B}\bar{\beta} _{A} \right)  \right]
\nonumber
\\
&+&\left[-\frac{1}{2}\left(\bar{\beta}^{C}\bar{\beta}_{C}-\bar{\alpha} \right)
\partial _{r}\bar{\beta}_{A}+\frac{1}{2}\hat{D}_{A}\bar{\alpha}-\frac{1}{2}\bar{\beta}^{B}
\left(\partial _{u}\mu _{AB}+\hat{D}_{A}\bar{\beta}_{B}-\hat{D}_{B}\bar{\beta} _{A} \right)  \right]
\left[-\frac{1}{2}\partial _{r}\bar{\beta} _{A} \right]
\nonumber
\\
&+&\left[-\frac{1}{2}\left\lbrace \partial _{u}\mu _{BC}+\left(\bar{\beta}^{M}\bar{\beta}_{M}
-\bar{\alpha} \right)\partial _{r}\mu _{BC}\right\rbrace 
+\frac{1}{2}\left(\hat{D}_{C}\bar{\beta} _{B}+\hat{D}_{B}\bar{\beta} _{C} \right) \right]
\nonumber
\\
&\times&\left[\frac{1}{2}\bar{\beta}^{C}\partial _{r}\bar{\beta}_{A}
+\frac{1}{2}\mu ^{CB}\partial _{u}\mu _{BA}+ \frac{1}{2}\mu ^{CB}
\left(\hat{D}_{B}\bar{\beta}_{A}-\hat{D}_{A}\bar{\beta} _{B}\right) \right]-\left(B\leftrightarrow u\right)\Big)
\nonumber
\\
&=&\frac{1}{2}\partial _{u}^{2}\mu _{AB}-\frac{1}{4}\partial _{u}\mu _{BC}\mu ^{CM}\partial _{u}\mu _{MA}
+\frac{1}{2}\alpha \partial _{u}\mu _{AB}
\end{eqnarray}
and
\begin{eqnarray}\label{Paper2:Eqa12}
R^{r}_{~PQR}&=&\partial _{Q}\Gamma ^{r}_{PR}-\partial _{R}\Gamma ^{r}_{PQ}+\Gamma ^{r}_{mQ}\Gamma ^{m}_{PR}
-\Gamma ^{r}_{mR}\Gamma ^{m}_{PQ}
\nonumber
\\
&=&-\partial _{Q}
\left(\frac{1}{2}\left\lbrace \partial _{u}\mu _{PR}+\left(\bar{\beta}^{C}\bar{\beta}_{C}
-\bar{\alpha} \right)\partial _{r}\mu _{PR}\right\rbrace 
+\frac{1}{2}\left(\hat{D}_{P}\bar{\beta} _{R}+\hat{D}_{R}\bar{\beta} _{P} \right) \right)
\nonumber
\\
&+&\left(\frac{1}{2}\left(\bar{\beta}^{C}\bar{\beta}_{C}-\bar{\alpha} \right)
\partial _{r}\bar{\beta}_{Q}+\frac{1}{2}\hat{D}_{Q}\bar{\alpha}
-\frac{1}{2}\bar{\beta}^{B}\left(\partial _{u}\mu _{QB}+\hat{D}_{Q}\bar{\beta}_{B}
-\hat{D}_{B}\bar{\beta} _{Q} \right) \right)
\left(\frac{1}{2}\partial _{r}\mu _{PR} \right)
\nonumber
\\
&-&\left(\frac{1}{2}\left(\partial _{r}\bar{\beta}_{Q}
-\bar{\beta}^{C}\partial _{r}\mu _{CQ} \right) \right)
\left(\frac{1}{2}\left\lbrace \partial _{u}\mu _{PR}
+\left(\bar{\beta}^{C}\bar{\beta}_{C}-\bar{\alpha} \right)\partial _{r}\mu _{PR}\right\rbrace 
+\frac{1}{2}\left(\hat{D}_{P}\bar{\beta} _{R}+\hat{D}_{R}\bar{\beta} _{P} \right) \right)
\nonumber
\\
&-&\left(\frac{1}{2}\left\lbrace \partial _{u}\mu _{AQ}+\left(\bar{\beta}^{C}\bar{\beta}_{C}
-\bar{\alpha} \right)\partial _{r}\mu _{AQ}\right\rbrace 
+\frac{1}{2}\left(\hat{D}_{A}\bar{\beta} _{Q}+\hat{D}_{Q}\bar{\beta} _{A} \right) \right)
\left(\frac{1}{2}\bar{\beta}^{A}\partial _{r}\mu _{PR}+\hat{\Gamma}^{A}_{PR} \right)-(Q\leftrightarrow R)
\nonumber
\\
&=&-\frac{1}{2}\partial _{Q}\partial _{u}\mu _{PR}+\frac{1}{2}\partial _{R}\partial _{u}\mu _{PQ}
+\frac{1}{4}\beta _{Q}\partial _{u}\mu _{PR}
-\frac{1}{4}\beta _{R}\partial _{u}\mu _{PQ}
\nonumber
\\
&-&\frac{1}{2}\partial _{u}\mu _{AQ}\hat{\Gamma}^{A}_{PR}
+\frac{1}{2}\partial _{u}\mu _{AR}\hat{\Gamma}^{A}_{PQ}
\end{eqnarray}

\section{Identities Regarding Lie Variation of $P^{abcd}$}\label{Paper2:AppB}

In this section we shall derive some identities related to Lie variation of 
the entropy tensor, $P^{abcd}$. 
For that purpose we first consider Lie variation of the Lagrangian treated as a 
scalar function of the metric $g_{ab}$ and $R_{abcd}$ leading to
\begin{equation}\label{Paper2:Eqb01}
\pounds_{\xi}L\left(g_{ab},R_{ijkl}\right)=\xi ^{m}\nabla _{m}L\left(g_{ab},R_{ijkl}\right)
=\frac{\partial L}{\partial g_{ab}}\xi ^{m}\nabla _{m}g_{ab}+\frac{\partial L}{\partial R_{ijkl}}
\xi ^{m}\nabla _{m}R_{ijkl}=P^{ijkl}\xi ^{m}\nabla _{m}R_{ijkl}
\end{equation}
where we have used the fact that covariant derivative of metric 
tensor vanishes. Then for the Lagrangian which is homogeneous function of 
degree m we get
\begin{equation}\label{Paper2:Eqb02}
\pounds_{\xi}L=\xi ^{m}\nabla _{m}\left(\frac{1}{m}P^{ijkl}R_{ijkl}\right)
\end{equation}
Then using \eq{Paper2:Eqb01} we readily obtain
\begin{equation}\label{Paper2:Eqb03}
R_{abcd}\xi ^{m}\nabla _{m}P^{abcd}=(m-1)P^{abcd}\xi ^{m}\nabla _{m}R_{abcd}
\end{equation}
We also have the following relation:
\begin{eqnarray}\label{Paper2:Eqb04}
P^{ijkl}\pounds_{\xi}R_{ijkl}&=&P^{ijkl}\left(\xi ^{m}\nabla _{m} R_{ijkl}+R_{ajkl}\nabla _{i}\xi^{a}
+R_{iakl}\nabla _{j}\xi^{a}+R_{ijal}\nabla _{k}\xi^{a}+R_{ijka}\nabla _{l}\xi^{a}\right)
\nonumber
\\
&=&P^{ijkl}\xi ^{m}\nabla _{m} R_{ijkl}+4\nabla _{i}\xi _{m}R^{i}_{~jkl}P^{mjkl}
\nonumber
\\
&=&P^{ijkl}\xi ^{m}\nabla _{m} R_{ijkl}+4\nabla _{i}\xi _{m}\mathcal{R}^{im}
\end{eqnarray}
Again we can also write, $m\pounds_{\xi}L=P^{ijkl}\pounds_{\xi}R_{ijkl}+R_{ijkl}
\pounds_{\xi}P^{ijkl}$. Then we obtain:
\begin{eqnarray}\label{Paper2:Eqb05}
R_{ijkl}\pounds_{\xi}P^{ijkl}&=&m\pounds_{\xi}L-P^{ijkl}\pounds_{\xi}R_{ijkl}
\nonumber
\\
&=&m\pounds_{\xi}L-P^{ijkl}\xi ^{m}\nabla _{m} R_{ijkl}-4\nabla _{i}\xi _{m}\mathcal{R}^{im}
\nonumber
\\
&=&(m-1)P^{abcd}\xi ^{m}\nabla _{m}R_{abcd}-4\nabla _{i}\xi _{m}\mathcal{R}^{im}
\end{eqnarray}
This equation can also be casted in a different form as
\begin{equation}\label{Paper2:Eqb06}
R_{abcd}\left(\pounds_{\xi}P^{abcd}-\xi ^{m}\nabla _{m}P^{abcd}\right)=-4\nabla _{i}\xi _{m}\mathcal{R}^{im}
\end{equation}
Now we can rewrite the metric as a function of $g_{ab}$ and $R^{a}_{~bcd}$, in which case the Lie 
variation leads to
\begin{equation}\label{Paper2:Eqb07}
\pounds_{\xi}L\left(g_{ij},R^{a}_{~bcd}\right)=P_{i}^{~jkl}\xi ^{m}\nabla _{m}R^{i}_{~jkl}
\end{equation}
With the Lagrangian as homogeneous function of curvature tensor to mth order leads to
\begin{equation}\label{Paper2:Eqb08}
R^{a}_{~bcd}\pounds _{\xi}P_{a}^{~bcd}=(m-1)P_{i}^{~jkl}\xi ^{m}\nabla _{m}R^{i}_{jkl}
\end{equation}
Then we arrive at the following identity
\begin{equation}\label{Paper2:Eqb09}
P^{abcd}\pounds _{\xi}\left(g_{am}R^{m}_{~bcd}\right)=P_{a}^{~bcd}\xi ^{m}\nabla _{m}R^{a}_{~bcd}
+4\nabla _{i}\xi _{m}\mathcal{R}^{im}
\end{equation}
or
\begin{eqnarray}\label{Paper2:Eqb10}
P_{a}^{~bcd}\pounds_{\xi}R^{a}_{~bcd}&=&P_{a}^{~bcd}\xi ^{m}\nabla _{m}R^{a}_{~bcd}
+4\nabla _{i}\xi _{m}\mathcal{R}^{im}-\mathcal{R}^{am}\mathcal{L}_{\xi}g_{am}
\nonumber
\\
&=&P_{a}^{~bcd}\xi ^{m}\nabla _{m}R^{a}_{~bcd}
+2\nabla _{i}\xi _{m}\mathcal{R}^{im}
\end{eqnarray}
This leads to the following relation:
\begin{equation}\label{Paper2:Eqb11}
R^{a}_{~bcd}\left(\pounds_{\xi}P_{a}^{~bcd}-\xi ^{m}\nabla _{m}P_{a}^{~bcd}\right)
=-2\nabla _{i}\xi _{m}\mathcal{R}^{im}
\end{equation}
If we proceed along the same lines we readily obtain another 
such relation given as:
\begin{equation}\label{Paper2:Eqb12}
R^{ij}_{kl}\left(\pounds_{\xi}P^{kl}_{ij}-\xi ^{m}\nabla _{m}P^{kl}_{ij}\right)=0
\end{equation}
These relations illustrate the Lie variation of $P^{abcd}$ when contracted with the 
curvature tensor.


\end{document}